\documentclass[12pt]{article}
\usepackage{amsmath}
\usepackage{amssymb,graphicx,amsthm}
\usepackage{hyperref}
\numberwithin{equation}{section}
\newcommand{\der}{\partial}
\newcommand{\e}{\mathrm{e}}
\newcommand{\ii}{\mathrm{i}}
\newcommand{\dd}{\mathrm{d}}

\newcommand{\eps}{\varepsilon}
\newcommand{\RR}{\mathbb R}

\newcommand{\tend}{\rightarrow}
\newcommand{\then}{\Rightarrow}
\newcommand{\Tr}{\mathrm{Tr}}
\newcommand{\sla}{\!\!\!\!\slash}
\newcommand{\Ree}{\Re\mathrm{e}\,}
\newcommand{\Imm}{\Im\mathrm{m}\,}

\newcommand{\veps}[2]{{\varepsilon}^{#1}_{\phantom{#1}{#2}}}

\newcommand{\Sigmahat}[2]{{\Sigma}^{\hat #1}_{\phantom{#1}\hat {#2}}}
\newcommand{\Sigmatilde}[2]{{\Sigma}^{\tilde #1}_{\phantom{#1}{\tilde #2}}}
\newcommand{\sth}{\sigma_{\hat 3}}

\title{1/8 BPS States in AdS/CFT}
\date{}
\begin{document}
\pagestyle{empty}
\begin{center}
{\Large\bf 1/8 BPS States in AdS/CFT}
\\[2.1em]

\bigskip

{\large Edi Gava$^{a,b,c}$, Giuseppe Milanesi $^{c}$,\\
 K.S. Narain$^a$ and Martin O'Loughlin$^a$}\\

\null

\noindent 
  {\it $^a$ High Energy Section,
The Abdus Salam International Centre for Theoretical Physics,
Strada Costiera 11, 34014 Trieste, Italy}
\\[2.1ex]
{\it $^b$ Istituto Nazionale di Fisica Nucleare, sez. di Trieste}\\[2.1ex]
{\it $^c$ Scuola Internazionale Superiore di Studi
Avanzati, \\
Via Beirut 2-4, 34014 Trieste, Italy}
\\
%[2.1ex]
 \vfill

\end{center}

\begin{abstract}
We study a class of exact supersymmetric solutions of type IIB Supergravity. 
They have an $SO(4)\times SU(2) \times U(1)$ isometry and preserve generically 
4 of the 32 supersymmetries of the theory. Asymptotically $AdS_5 \times S^5$ 
solutions in this class are dual to 1/8 BPS chiral operators which
preserve the same symmetries in the $\mathcal N=4$ SYM theory. They are parametrized by a 
set of four functions that satisfy certain differential equations. 
We analyze the solutions to these equations in a large radius asymptotic expansion:
they carry charges with respect to two $U(1)$  KK gauge fields
and their mass saturates the expected BPS bound. 
\end{abstract}
\vfill 
\rightline{SISSA 63/2006/EP}
\rightline{IC/2006/116}
\newpage
\pagestyle{plain}
\section{Introduction}
One of the first steps in understanding the $AdS / CFT$ 
correspondence is to set up a precise dictionary  between the states 
of the  theories on the two sides of the correspondence. 
It is well known that the parameters of the $\mathcal N = 4$ $SU(N)$ SYM, 
namely $\lambda \equiv g_{YM}^2 N$ and $N$ should be identified 
with the parameters of type IIB String Theory on 
$AdS_5\times S^5$, namely $L_{AdS},\ell_s,g_s$, as 
\begin{equation}
	\frac{L_{AdS}^2}{4\pi \ell_s^2}=\left(\frac{\lambda}
{4\pi}\right)^{\frac 12}\qquad g_s = \frac \lambda N\,.
\end{equation}
Type IIB Supergravity is a good approximation of String 
Theory at low energies compared to the string scale and small string coupling. 
We may thus consider solutions to the supergravity equations 
of motion which are asymptotically $AdS_5 \times S^5$ as good 
candidates for dual of states in the CFT,  provided that 
$N\gg \lambda \gg 1$ and  any dimension four  curvature 
invariant of the solutions, $ \mathcal R_4$, satisfies
\begin{equation}
	\mathcal R_4  L_{AdS}^4 \ll \lambda \ll N\,.
\end{equation}

One may hope to be able to carry out this program in the full BPS 
sector of the respective
dual theories. A very beautiful and relatively simple construction 
of such a dictionary in the
half BPS sector has been performed in \cite{bubbling}(LLM). 
The authors considered  geometries dual to half BPS states in the CFT,  
associated to chiral primary operators which are obtained by taking 
traces of powers of the operator $Z\equiv Z_3\equiv 
\phi_5+\ii\phi_6$,  where $\{\phi_i\}_{1\leq i\leq 6}$ are 
the six adjoint scalars of the ${\cal N}=4$ $SU(N)$ gauge theory. 
In LLM, exact half BPS solutions to the supergravity 
equations of motion are derived by exploiting the 
$\RR \times SO(4)\times SO(4)$ bosonic symmetry of the problem. 
The complete geometry,  together with
the  self-dual 5-form, are determined by a single function 
$z$ which is defined on a three dimensional halfspace 
and which satisfies a linear elliptic differential equation: 
solutions are thus specified by boundary conditions at 
infinity and on the boundary plane.  
LLM were able to identify the boundary conditions 
giving rise to non-singular asymptotically $AdS_5 \times S^5$ geometries. 
The resulting space of classical solutions can be 
directly identified with the phase space of the dual states of the gauge 
theory in the free fermion picture \cite{Corley:2001zk,Berenstein:2004kk}.
The latter emerges after reducing the (single) scalar sector of 
the gauge theory on $\RR \times S^3$. 
These solutions represent the geometrical transition 
between probe giant gravitons or dual giant gravitons 
\cite{Suryanarayana:2004ig,McGreevy:2000cw,Grisaru:2000zn,Hashimoto:2000zp} 
and fully backreacted geometries. A giant
graviton is a classical D3-brane configuration
wrapping an $S^3 \subset S^5$ and
rotating along an equator of the $S^5$. A dual-giant graviton
is another half-BPS D3-brane configuration that wraps an $S^3 \subset AdS_5$. 

It is natural to ask how the above very precise correspondence
between geometry on the one hand, and features of the quantum mechanical 
states of the reduced gauge theory on the other, 
extends to cases with less supersymmetry.   
In the recent literature there have been various attempts  
in this direction:  for example,
in \cite{Liu:2004ru}   one quarter BPS geometries 
were found by assuming a non trivial axion-dilaton. 
This corresponds to putting smeared D7 branes in the background 
and thus to adding flavour to the gauge theory. A description 
of one eighth and one quarter BPS geometries in the language 
of five dimensional gauged supergravity has been given in 
\cite{Chong:2004ce}. The construction of a class of one quarter 
BPS solutions directly in type IIB appeared in 
\cite{Donos:2006iy,Donos:2006ms}. 
Another interesting related work is presented in \cite{Berenstein:2005aa}.
This problem was also approached in the probe
approximation, where the backreaction on the geometry
is neglected: D3 branes can wrap more complex three dimensional 
surfaces in $S^5$ and give rise to 
giant gravitons with fewer supersymmetries \cite{Mikhailov:2000ya}. 
In \cite{Mandal:2006tk} the authors have been able 
to count such states. The quantization of their classical 
phase space has been performed in \cite{Biswas:2006tj}. Other works that present interestng connection with ours can be found in \cite{Kim:2005ez,Kim:2006qu,Lunin:2006xr}.

In this paper we address the problem of finding BPS 
supergravity solutions which represent the fully backreacted 
geometry of a class of 1/8 BPS giant gravitons. 
Our solutions correspond to gauge theory states 
associated to linear combinations of composite operators
\begin{equation}
	\mathcal O (q,r) = \Tr (Z_1^{q}) \Tr (Z_2^{q}) \Tr (Z_3^r)+\cdots\,.
\end{equation}
where $Z_1,Z_2$ and $Z_3$ are the three complex scalars of the 
$\mathcal N =4$ CFT. The dots signify other terms with suitable (anti)-symmetrization and trace structures, which have all a total of $q$ $Z_1$ and $Z_2$ fields and $r$ $Z_3$ fields.  They are chosen such that $\mathcal O(q,r)$ are chiral primary operators which are invariant under the $SU(2)_L$ subgroup of the $SU(2)_L \times SU(2)_R$ acting on $Z_1,Z_2$. We consider linear combinations of $\mathcal O(q,r)$ which have all the same value of $q$ but may have different values of $r$.\\
The lowest mode $\mathcal O(q,r)$ in the expansion on spherical harmonics on $S^3$ saturates the BPS bound:
\begin{equation}
	\Delta = 2q+r\,,
\end{equation}
where $\Delta$ is the conformal dimension of the operator.
The total amount of bosonic symmetry preserved by the corresponding states is thus given by $SO(4)_{KK}\times SU(2)_L \times U(1)_R$.
 Consequently, we start from an 
Ansatz for the metric and the self-dual RR 5-form 
which preserves this amount of symmetry.  This implies, as for LLM,
that the resulting background will depend non-trivially 
on three coordinates (an additional
symmetry will be associated to the time coordinate, like in LLM).
We also require that the background possesses the required 
amount of supersymmetry by demanding that it possesses a Killing spinor. 
Applying techniques similar to those in 
\cite{bubbling,Gauntlett:2002sc,Gauntlett:2002nw,Gutowski:2003rg,Gauntlett:2004zh,Gauntlett:2004yd} 
we have been able to express 
the full solution in terms of four independent functions 
defined on a three dimensional half-space. As a result 
of certain Bianchi identities and integrability conditions, these four functions 
have to satisfy a system of nonlinear, 
coupled, elliptic differential equations. A unique solution 
to these equations is obtained once a set of boundary conditions 
at infinity and on the boundary plane is specified; boundary conditions 
should be chosen in such a way as to give non-singular geometries with
$AdS_5 \times S^5$ asymptotics.

We present here the boundary conditions that give 
rise to asymptotically non-singular  $AdS_5 \times S^5$ geometries. We
solve the equations asymptotically up to third order in a large radius
expansion. From this analysis we can extract the two 
dimensionless charges $Q$ and $J$ carried by the solution. 
These are the charges corresponding  to  two  out of the three  
$U(1)$ Cartan gauge fields arising from the KK reduction of 
IIB supergravity on $S^5$ to five dimensional
maximal gauged supergravity.  These charges in turn correspond to 
the $q$ and $r$ charges of the gauge theory side. 
Moreover, we verify that our solutions saturate the expected BPS bound:
\begin{equation}
	M = \frac{\pi L_{AdS}^2}{4 G_5}(|J|+2| Q|)\,.
\end{equation}

Unfortunately, a more exhaustive analysis of such boundary 
conditions is quite difficult due to the complexity (non linearity)  
of the differential equations. In other words we do not know
which of the boundary conditions give rise to globally non-singular
backgrounds. We will comment on this issue  in the  conclusions.
The paper is organized as follows. In Section 2 we present the gauge theory 
description of the 1/8 BPS states that we wish to study. In Section 3 we show how 
the 1/8 supersymmetry constrains the components of the metric and 5-form and 
we reduce these constraints to four differential equations on four scalar functions.  
In Section 4 we present the large radius asymptotic analysis. 
Appendix A sets our conventions. In Appendix B the complete derivation of the 
results presented in Section 3 is 
given and in Appendix C we make some observations on the formal 
tools used to facilitate the analysis. 
Due to the complexity of the equations involved, we performed 
the complete analysis by means of the software Mathematica. All the derivations 
that are not described in full detail in the text were 
obtained with the help of such software.

\section{Gauge theory analysis}\label{gaugetheory}
Chiral primary operators of the $\mathcal N = 4$
superconformal algebra $SU(2,2|4)$ were classified
in  \cite{Dobrev:1985qv,Andrianopoli:1998ut,Andrianopoli:1999vr} and are characterized
by the number of Poincare' supersymmetries they preserve.
They can preserve $\frac{1}{2}$,  $\frac{1}{4}$ or $\frac{1}{8}$
of the 16 Poincare' supersymmetries. We will be interested in
operators that are composites of the six adjoint scalars $\phi_i$
of the 
$\mathcal N = 4$ $SU(N)$ gauge theory. The simplest class
is that of $\frac{1}{2}$-BPS operators the elements of which are characterized by the $SO(6)$
R-symmetry representations given by Young tableaux with a single
row of length $p$, i.e. traceless, symmetric $SO(6)$ rank $p$ 
tensors, or in Dynkin notation, the $[0,p,0]$ of $SU(4)$. In this case 
the conformal dimension $\Delta$ is equal to $p$. 
The highest weight of this representation can be obtained
by using one of the  complex adjoint scalars, $Z\equiv Z_3\equiv \phi_5+i\phi_6$, 
which has charge 1 with respect to
the $SO(2)$ generator, $J_3\equiv L_{5,6}$\footnote
{Here $J_i=L_{2i-1, 2i}$ in terms of the standard generators
of $SO(6)$} , 
in $SO(6)$. One can construct from $Z$ the multitrace
composite $SU(N)$ singlet operators with $\Delta=p$. These operators 
therefore preserve an $SO(4)\subset SO(6)$  times an $SO(4)\subset SO(4,2)$, 
since the only modes satisfying the relation $\Delta = p$ are the $S^3$ scalars.
This $SO(4)\times SO(4)$ symmetry has been used in LLM as the isometry
of the supergravity background and it is the key for their (relatively)
simple and beautiful solution. 
The lower supersymmetry cases are again best described in terms
of $SO(6)$ Young tableaux: the $\frac{1}{4}$ and $\frac{1}{8}$ 
cases correspond to  tableaux with two rows (of lengths $p,q$,
$p\geq q$)
and three rows (of lengths $p,q,r$, $p\geq q \geq r$) respectively. The  
conformal dimensions saturate the bounds $p+q$ and 
$p+q+r$ respectively.
Again, in discussing highest weight states it is 
convenient to use the three complex scalars  $Z_1=\phi_1+i\phi_2$, 
$Z_2=\phi_3+i\phi_4$ and $Z_3$,
which have charges (1,0,0), (0,1,0) and (0,0,1) with
respect to the three Cartan generators $(J_1,J_2,J_3)$  
$SU(4)$ respectively. Highest weight states saturate the BPS bound 
$\Delta=J_1+J_2=p+q$ and $\Delta=J_1+J_2+J_3=p+q+r$ 
in the  $\frac{1}{4}$ and $\frac{1}{8}$ case, respectively.

This is summarized in the following table:
\begin{center}
$\begin{array}{c|c|c}
 p = q = 0, \,r \neq 0 & p,q \neq 0,\, r = 0 & p,q,r\neq 0\\
\hline
1/2 \, BPS & 1/4 \, BPS & 1/8 \,BPS\\
\end{array}$
\end{center}
Let us consider the $\frac{1}{8}$ case: given 
the three complex scalars $Z_1,Z_2,Z_3$ of the  
$\mathcal N = 4$ $SU(N)$ Super Yang Mills theory 
one can construct a basis of gauge invariant, local, 
composite operators in the 
$[p,q,r]$ of the $R$-symmetry group  $SU(4)$ as \cite{Ryzhov:2001bp}
\begin{equation}
	\Tr (Z_1^{p}) \Tr (Z_2^{q}) \Tr (Z_3^r)+\cdots\,.
\end{equation}
where the dots mean suitable (anti)-symmetrization and trace structure that projects to the chiral primaries in the $(p,q,r)$ representation of $SU(4)$.

%By taking linear combinations of these operators, 
%one can construct chiral primaries BPS operators 
%which preserve a fraction of the supersimmetries a
%ccording to the following table
%\begin{center}
%$\begin{array}{c|c|c}
% p = q = 0 \,r \neq 0 & p,q \neq 0,\, r = 0 & p,q,r\neq 0\\
%\hline
%1/2 \, BPS & 1/4 \, BPS & 1/8 \,BPS\\
%\end{array}$
%\end{center}
%Given such a chiral primary $\Phi(x)$, it satisfies
%\begin{equation}
%	\big[ \Delta - (p+q+r) \big] \Phi (0) = 0
%end{equation}

We are interested in constructing duals of the 
states corresponding to such operators. However,
generic operators of this type break fully the non-abelian $SO(6)$
R-symmetry, up to possible $U(1)$ factors which act by an overall
phase on them.  However, if
\begin{equation}
	p = q\,.
\end{equation}
we can construct operators which are invariant under the $SU(2)_L$ 
of the $SO(4)= SU(2)_L\times SU(2)_R$ which rotates the four real scalars
\begin{equation}
	\begin{pmatrix}
	\phi_1\\\phi_2\\\phi_3\\\phi_4
	\end{pmatrix} =
\begin{pmatrix}
	\Ree Z_1 \\ \Imm Z_1 \\ \Ree Z_2 \\ \Imm Z_2
\end{pmatrix}\,.
\end{equation}
This is best seen by observing that $SU(2)_L$  and $SU(2)_R$ 
act as left and right multiplication respectively on:
\[ \left(\begin{array}{rr}
Z_1 & -\bar{Z_2} \\ Z_2 & \bar{Z_1}
\end{array}\right)\]
Therefore $Z_1$ and $Z_2$ transform as a doublet of $SU(2)_L$,
whereas they have the same charge under $J^3_R = \frac{J_1+J_2}{2}$.
The operators with $p=q$ are clearly singlets of  $SU(2)_L$, and
they acquire an overall phase under $J^3_R$. They satisfy the relation
\begin{equation}
	\Delta = 2 q + r\,.
\end{equation}

The bosonic symmetry preserved by these states is:
\begin{equation}
	\RR_{BPS}\times \big(SU(2)_L \times 
U(1)_R\big)_{R- \textrm{charge}}\times SO(4)_{KK}
\end{equation}
where the first $\RR$ corresponds to the transformations generated by
\begin{equation}
	D^\prime \equiv D - 2 J^3_R - J\,, 
\end{equation}
with $J=J_3$ acting on $Z$, $D$ is the dilatation operator
and the last $SO(4)$ factor represents the fact that 
we are considering $s$-wave modes on $S^3$ in the reduction
of SYM theory on $\RR\times S^3$ \cite{Witten:1998qj}.
These are the symmetries that will motivate the Ansatz for the metric and
five-form on the supergravity side: we will keep a round 3-sphere 
with the $SO(4)$ isometry corresponding to the $SO(4)$ above. Another
$S^3$ (related to the $SO(4)$ R-symmetry of the $\frac {1}{2}$ BPS case) 
which is in the $S^5$ of the 
$AdS_5\times S^5$ background, will be squashed with isometry group reduced to 
$SU(2)_L \times U(1)_R$. 

It will be useful for the subsequent analysis of the Killing
spinor equation on the supergravity side, to understand the quantum numbers
of the preserved supersymmetries. In an $\mathcal N=1$ and 
$SU(3)\times U(1)\subset SU(4)$ notation,
the supersymmetry variations of the complex
scalars $Z_i$ are:
\begin{equation}
\delta{Z_i}=\xi_i\lambda+\xi \psi_i+\epsilon_{ijk}\bar{\xi}^j\bar{\psi}^k,
\label{susy}
\end{equation}  
Here the two-component spinors $\lambda$ and $\psi_i$ 
are the gaugino and the chiral matter fermions,
while $\xi$ and $\xi_i$ are the supersymmetry parameters.
They are in the ${\bf 1}_{3/2}$ and ${\bf 3}_{-1/2}$ of $SU(3)_{U(1)}$ 
respectively. More precisely the 
Cartan charges of $\lambda, \xi$ are $(\frac{1}{2},\frac{1}{2},\frac{1}{2})$,
and those of $\psi_1, \xi_1$ are $(\frac{1}{2},-\frac{1}{2},-\frac{1}{2})$,
and similarly for $\psi_{2,3},\xi_{2,3}$. 
 From \eqref{susy} it is clear that the highest weight 
$\frac{1}{8}$ BPS operators are invariant under the supersymmetry
corresponding to $\bar{\xi}$. As for the $SU(2)_L\times SU(2)_R=
SO(4)\subset SU(3)$ quantum numbers, the roots of $SU(2)_L$ are $(\pm 1,
\mp 1, 0)$ and those of $SU(2)_R$ are  $(\pm 1,\pm 1, 0)$. Therefore
the preserved supersymmetry parameter $\bar\xi$, whose charges are
$(-\frac{1}{2},-\frac{1}{2},-\frac{1}{2})$, is a singlet of the 
unbroken $SU(2)_L$ and lowest weight with respect to the broken $SU(2)_R$.

\section{Generic Solutions}
We are looking for supergravity solutions dual to BPS states constructed from linear combinations of the operators
\begin{equation}
	\mathcal O(q,r)=\Tr \big( Z_1^q\big) \Tr  \big(Z_2^q\big)\Tr \big( Z_3^{r} \big)+\cdots
\end{equation}
for constant $q$, where the meaning of the dots has been explained in the previous two sections.
The geometries will thus be invariant under $SU(2)_L \times SO(4)_{KK}$ as defined in the previous section and invariant but charged under the remaining $U(1)_R$. The extra non-compact time-like 
symmetry ($\RR_{BPS}$ of the previous section) is associated to invariance under 
the transformations generated by $D^\prime$ in the 
gauge theory and will emerge naturally in our construction\footnote{See Appendix B for details.}.

The most generic Ansatz consistent with these symmetries is given by
\begin{equation}
	\dd s^2 = g_{\mu\nu} \dd x^\mu \dd x^\nu + \rho_1^2 \big[(\sigma^{\hat 1})^2 + (\sigma^{\hat 2})^2\big] +\rho_3^2(\sigma^{\hat 3} - A_\mu \dd x^\mu)^2 + \tilde \rho^2 \dd \tilde \Omega_3^2\,.
\end{equation}
where $\rho_1$, $\rho_3$, $\tilde{\rho}$, $A_{\mu}$ and $g_{\mu \nu}$ are functions of the four
coordinates $x^{\mu}$.
The space is a fibration of a squashed 3-sphere (on which the $SU(2)$ left-invariant 1-forms $\sigma^{\hat a}$ are defined) and a round 3-sphere (on which the $SU(2)$ left-invariant 1-forms $\sigma^{\tilde a}$ are defined) over a four dimensional manifold.\\
The left invariant 1-forms are given by
\begin{equation}
\begin{array}{ll}
\sigma^{\hat 1}= -\frac 12 (\cos \hat\psi\, d\hat\theta + \sin \hat\psi\, \sin\hat \theta\, d\hat \phi) & \sigma^{\tilde 1}= -\frac12(\cos \tilde\psi\, d\tilde\theta + \sin \tilde\psi\, \sin \tilde\theta\, d\tilde\phi)\\
\sigma^{\hat 2}= -\frac12(-\sin\hat \psi\, d\hat\theta + \cos\hat \psi\, \sin\hat \theta\, d\hat \phi )& \sigma^{\tilde 2}= -\frac12(-\sin \tilde\psi\, d\tilde\theta + \cos \tilde\psi\, \sin\tilde \theta\, d\tilde\phi) \\
\sigma^{\hat 3}= -\frac12(d\hat\psi + \cos \hat\theta\, d \hat\phi) & \sigma^{\tilde 3}= -\frac12(d\tilde\psi + \cos \tilde\theta\, d \tilde\phi) 
\label{sigmas}
\end{array}
\end{equation}
and satisfy the relations
\begin{equation}
\begin{split}
	\dd \sigma^{\hat i } = \epsilon_{\hat i \hat j \hat k} \sigma^{\hat j }\wedge \sigma^{\hat k}\\
\dd \sigma^{\tilde i} = \epsilon_{\tilde i \tilde j \tilde k} \sigma^{\tilde j}\wedge \sigma^{\tilde k}\,.
\end{split}
\end{equation}
With this normalization the metric on the unit radius round three sphere is given by
\begin{equation}
	d \Omega_3^2 = (\sigma^1)^2+(\sigma^2)^2+(\sigma^3)^2\,,
\end{equation}
with $\sigma^a$ being either $\sigma^{\hat a}$ or $\sigma^{\tilde a}$.\\
We choose our ``d-bein'' to be
\begin{align}
e^m =& \veps{m}{\mu}\dd x^\mu\\
e^{\hat a} =& \begin{cases}
	\rho_1 \sigma^{\hat a} & a=1,2\\
	\rho_3(\sigma^{\hat 3} -A_\mu\dd x^\mu) & a=3
      \end{cases}\\
e^{\tilde a} =& \tilde \rho \sigma^{\tilde{a}}
\end{align}
Since we are looking for the geometric dual to operators which involve only scalar fields 
in the gauge theory, the only possible non-zero Ramond-Ramond field strength is the 
five form $F_{(5)}$ 
and the dilaton is assumed to be constant. The most generic Ansatz for the five form which is 
invariant under the given symmetries is:
\begin{multline}
	F_{(5)} =  2 \left( \tilde G_{mn} e^m\wedge e^n + \tilde V_m e^m\wedge e^{\hat 3} + \tilde g e^{\hat 1}\wedge e^{\hat 2} \right)\wedge \tilde\rho^3\dd \tilde \Omega_3+\\
	2\left(- G_{pq}e^p\wedge e^q \wedge e^{\hat 1}\wedge e^{\hat 2}\wedge  e^{\hat 3} +\star_4 \tilde V\wedge  e^{\hat 1}\wedge e^{\hat 2}- \star_4 \tilde g \wedge e^{\hat 3}\right) \,,
\end{multline}
where
\begin{gather}
	G_{mn} = \frac12 \epsilon_{mnpq} \tilde G^{mn}\\
	\star_4 \tilde V = \frac{1}{3!}\epsilon_{mnpq}\tilde V^m e^n\wedge e^p\wedge e^q\\
	\star_4 \tilde g  = \tilde g e^0\wedge e^1\wedge e^2\wedge e^3
\end{gather}
The Bianchi identity $\dd F_{(5)} = 0$ implies:
\begin{gather}
	\dd \big(\tilde G \tilde \rho^3 - \tilde V \wedge A \rho_3\tilde\rho^3\big) = 0\label{bianchiGVA}\\
	\tilde V = \frac 12 \frac{1}{\rho_3 \tilde\rho^3}\dd (\tilde{g} \rho_1^2 \tilde\rho^3)\label{bianchitildeV}\\
	\dd \big( G \rho_1^2 \rho_3\big) = 0\label{bianchiG}\\
	\dd \big( G \rho_1^2 \rho_3 \wedge A + \star_4 \tilde V \big)-2\star_4 \tilde g = 0\,.\label{lastbianchi}
\end{gather}
Since we are looking for the dual of BPS states, the background should preserve a fraction 
of the supersymmetry and so there should exist a supersymmetry parameter $\psi$ such that the 
gravitino variation vanishes:
\begin{equation}
	\delta \chi_M = \nabla_M \psi + \frac{\ii}{480}F_{M_1 M_2 M_3 M_4 M_5} \Gamma^{M_1 M_2 M_3 M_4 M_5}\Gamma_M \psi = 0 \,.
\end{equation}
The Bianchi identity and the existence of the spinor $\psi$ are sufficient for our supergravity 
background to satisfy the full equations of motion of type IIB Supergravity.\\
The existence of the spinor $\psi$ is also sufficient to express the complete solution in the following form:
\begin{multline}
\dd s^2 = - h^{-2} (\dd t + V_i \dd x^i)^2 + h^2 \frac{\rho_1^2}{\rho_3^2}(T^2\delta_{ij}\dd x^i\dd x^j + \dd y^2)+\tilde\rho^2\dd\tilde\Omega_3^2+ \\
+ \rho_1^2 \big(\hat\sigma_1^2 + \hat\sigma_2^2\big) +\rho_3^2(\hat\sigma_3 - A_t \dd t - A_i \dd x^i)^2
\end{multline}
where the coordinate $y$ is the product of two radii
\begin{equation}
y=\rho_1 \tilde \rho > 0\,,
\end{equation}
and the function $h$ is given by
\begin{equation}
h^{-2} = \tilde{\rho}^2 + \rho_3^2(1 + A_t)^2\,.
\end{equation}
The vector $\partial_t$ is the Killing vector which generates the extra non-compact timelike $U(1)$ and thus all the entries of the metric depend only on $(x^1,x^2,y)$, where $y$ is constrained to be positive. They can be expressed in terms of four independent functions:
\begin{equation*}
	m,n,p,T
\end{equation*}
as follows:
\begin{equation}	
\begin{array}{lll}
 \rho_1^4 =  \frac{m p+ n^2}{m} y^4& \rho_3^4 = \frac{p^2}{m(mp+n^2)}&\tilde\rho^4 = \frac{m}{mp+n^2}\\ 
  h^{4} = \frac{ m p^2}{mp+n^2}& A_t = \frac{n-p}{p} & A_i = A_t V_i -\frac12 \epsilon_{ij}\der_j \ln T
\end{array}
\end{equation}
and
\begin{gather}
	\dd V = -y \star_3 [ \dd n + (n D +2y m (n-p) + 2n/y)\dd y]\label{firstdV}\\
	\der_y \ln T = D \\
	 D\equiv 2y(m+n-1/y^2)\label{firstdT}\,,
\end{gather}
where $\star_3$ indicates the Hodge dual in the three dimensional diagonal metric
\begin{equation}
	\dd s_3^2 = T^2\delta_{ij}\dd x^i\dd x^j + \dd y^2\,.
\end{equation}
The various four-dimensional forms from which the 5-form field strength is constructed are
\begin{gather}
\tilde g= \frac{1}{4 \tilde\rho}\left[1-\frac{\rho_3^2}{\rho_1^2}(1+A_t)\right]\\
	\tilde V = \frac 12 \frac{1}{\rho_3 \tilde\rho^3}\dd (\tilde{g} \rho_1^2 \tilde\rho^3) \\
G \rho_1^2 \rho_3 = \dd B_t \wedge (\dd t + V_i\dd x^i) + B_t \dd V + \dd \hat B\\
\tilde G \tilde \rho^3 = \frac12 g \rho_1^2 \tilde\rho^3  \dd A + \dd \tilde B_t \wedge (\dd t + V_i\dd x^i)+\tilde B_t \dd V + \dd \hat{\tilde B}\,,
\end{gather}
where
\begin{equation}
\begin{split}
 & \tilde B_t = -\frac{1}{16} y^2\, \frac{n-1/y^2}{p}\\
& \dd \hat{\tilde B} = -\frac1{16} y^3\star_3 [\dd m + 2 m D\,\dd y]\\
& B_t = -\frac 1 {16} y^2\,\frac{n}{m}\\
& \dd \hat B = \frac 1 {16}y^3\star_3  [\dd p + 4yn(p-n)\dd y]\,.\\
\end{split}
\end{equation}
\subsubsection*{Differential equations}
The Bianchi identities and the integrability condition for the equation \eqref{firstdV} give

\begin{equation}
\left\{ \begin{array}{l}
\dd \dd V = 0\\
\dd\dd \hat{\tilde B}=0\\
\dd\dd \hat B =0
\end{array}
\right.\,.
\end{equation}
These three conditions together with \eqref{firstdT} give a system of nonlinear coupled elliptic differential equations
\begin{equation}
\label{equations}
\begin{split}
&y^3 (\der_1^2+\der_2^2) n + \der_y \left(y^3 T^2 \der_y n\right) + y^2\der_y\big[ T^2 \big(y D n +2y^2m(n-p)\big)\big]+4 y^2 D T^2 n=0\\
&y^3 (\der_1^2+\der_2^2) m + \der_y \left(y^3 T^2 \der_y m\right) + \der_y \left( y^3 T^2 2m D\right)=0\\
&y^3 (\der_1^2+\der_2^2) p + \der_y \left(y^3 T^2 \der_y p\right) + \der_y \big[ y^3 T^2 4n y(n-p)\big]=0\\
&\der_y \ln T = D\,.
\end{split}
\end{equation}
A solution to these equations is determined by a set of boundary conditions at 
infinity (large values of $y,x^i$) and on the plane $y=0$; they should 
be chosen in such a way as to give a non-singular geometry  asymptotic to  $AdS_5\times S^5$.
Due to the non-linearity of the equations the relationship between  
boundary conditions and non-singular solutions with $AdS_5\times S^5$ asymptotics 
is difficult to control.
This set of boundary conditions may be regarded as a parametrisation of the space of solutions 
to our problem.

\subsubsection*{The LLM limit}

The LLM solutions are clearly a subset of ours. They are specified by the additional constraints,
\begin{equation}
	n=p=\frac 1{y^2} -m = \frac{1/2-z}{y^2}\quad T=1\,.
\end{equation}
In this case we have
\begin{equation}
	D=0\quad\rho_1=\rho_3=\rho\quad A_t = 0\quad T=1
\end{equation}
and the three second order equations collapse to one single linear equation
\begin{equation}
y^3 (\der_1^2+\der_2^2) n + \der_y \left(y^3 T^2 \der_y n\right)\,.
\end{equation}
As this 
equation is linear it has been possible to completely identify the boundary conditions at $y=0$ 
and at infinity that give rise to regular asymptotically $AdS_5\times S^5$ 
geometries\cite{bubbling,Milanesi:2005tp}. This set of boundary conditions can be directly 
identified with the classical phase space of the dual states in the free fermion picture.

\section{Asymptotics and Charges}
In this section we discuss asymptotic solutions to the differential equations of the 
previous section wich give $AdS_5 \times S^5$ asymptotics \footnote{A study of more general boundary conditions at $y=0$ will be presented in \cite{future}.}. We solve the equations to third order in an expansion for large values 
of $y,x^1,x^2$. 

We can identify the boundary conditions at infinity by comparing the leading order of this 
expansion to the same order of LLM, requiring in particular 
$AdS_5\times S^5$ asymptotics. 
The first corrections to the $AdS_5\times S^5$ geometry capture the global $U(1)$ charges 
under the gauge fields arising in the Kaluza Klein reduction of IIB supergravity over $S^5$. 
We will show that the solutions support 
non-vanishing fluxes for the the KK gauge fields associated to two of the three Cartan generators of the $SO(6)$ isometry of $S^5$. 
In the dual gauge theory picture these generators map to the $R$-symmetry generators 
$L_{5,6}$ and $L_{1,2}+L_{3,4}$.

It is not hard to see that the following expressions for our functions
\begin{equation}\label{asympexp}
\begin{split}
	&m \sim \frac 1{y^2} - \frac{p_1}{R^4}\\
	&n \sim \frac{p_1}{R^4}\\
	&p \sim \frac{p_1}{R^4}\\
	&T\sim 1
\end{split}
\end{equation}
satisfy the equations at leading order for large $R$, with $(R,\theta,\phi)$ polar coordinates in the $(x_1,x_2,y)$ space and $p_1$ is a constant parameter. We have also, to the same order,
\begin{equation}
	V_\phi \sim \frac{p_1 \cos^2\theta}{R^2} \qquad V_r \sim O \left(\frac 1 {R^4}\right)\,,
\end{equation}
with $r^2=x_1^2+x_2^2$, $r=R \cos \theta$ and $y=R\sin \theta$.\\
Defining
\begin{equation}
\begin{split}
&\tilde R = R/\sqrt{p_1}\\
&\tilde \phi = \phi - t
\end{split}	
\end{equation}
we get
\begin{equation}
	\dd s^2 = \sqrt p_1\left(- \tilde R^2  \dd t^2 + \frac{\dd \tilde R^2}{\tilde R^2} +  {\tilde R^2}\dd \tilde\Omega_3^2+\dd \theta^2  + \cos^2\theta \dd\tilde \phi^2 + \sin^2\theta\dd \hat\Omega_3^2\right)
\end{equation}
which is $AdS_5\times S^5$ in Poincare coordinates. The parameter $p_1$ and the radius $L$ of 
$AdS_5$ are related by 
\begin{equation}
L^2 = \sqrt{p_1}\,.
\end{equation}
We recall here the expression for the left-invariant one forms
\begin{equation}\label{LIsigmahat}
\begin{array}{l}
\sigma^{\hat 1}= -\frac 12 (\cos \hat\psi\, d\hat\theta + \sin \hat\psi\, \sin\hat \theta\, d\hat \phi)\\
\sigma^{\hat 2}= -\frac12(-\sin\hat \psi\, d\hat\theta + \cos\hat \psi\, \sin\hat \theta\, d\hat \phi )\\
\sigma^{\hat 3}= -\frac12(d\hat\psi + \cos \hat\theta\, d \hat\phi) \,.
\end{array}
\end{equation}
The metric on the unit radius round three sphere $\dd \hat \Omega_3$ is
\begin{equation}
	\dd \hat \Omega_3^2 = \big(\sigma^{\hat 1}\big)^2+\big(\sigma^{\hat 2}\big)^2+\big(\sigma^{\hat 3}\big)^2 = \frac 14 \big( \dd \hat \theta^2 + \dd \hat \phi^2 + \dd \hat \psi^2 + 2 \cos \hat \theta \dd \hat \psi \dd \hat \phi\big)\,.
\end{equation}
We can transform it into the more conventional form
\begin{equation}
	\dd \hat \Omega_3^2 = \dd \omega^2 + \cos^2 \omega\, \dd \phi_1^2 + \sin^2 \omega\, \dd \phi_2^2
\end{equation}
where
\begin{equation}\label{coordhat}
	\omega = \frac{\hat \theta } 2 \qquad \phi_1 = \frac{\hat\psi+\hat\phi}2 \qquad \phi_2 = \frac{\hat  \psi - \hat \phi}{2}\,.
\end{equation}

We will now consider the next two orders in the asymptotic expansion of our functions and solve the differential equations. For the sake of simplicity we will assume that $\der_\phi$ is also a Killing vector of our solutions. Despite this simplifying assumption, in general the solutions will still be charged under the corresponding KK gauge field. From the geometric point of view this means that the solutions are generically stationary. On the gauge theory side, this choice corresponds to looking for duals of linear combinations of states which have all the same $L_{56}$ charge\footnote{The analog of this choice in the LLM picture would be to consider solutions seeded by rotationally symmetric configurations of bubbles on the $y=0$ plane.} and are thus constructed from linear combinations of $\mathcal O (q,r)$ at fixed $q$ and $r$.

We thus assume the following expansion of our functions:
\begin{equation}\label{asympexpsub}
\begin{split}
	&m \sim \frac 1{y^2} - \frac{p_1}{R^4}+\frac{m_2(\theta)}{R^6}+\frac{m_3(\theta)}{R^8}\\
	&n \sim \frac{p_1}{R^4}+\frac{n_2(\theta)}{R^6}+\frac{n_3(\theta)}{R^8}\\
	&p \sim \frac{p_1}{R^4}+\frac{p_2(\theta)}{R^6}+\frac{p_3(\theta)}{R^8}\\
	&T\sim 1 + \frac{t_1(\theta)}{R^2}+ \frac{t_2(\theta)}{R^4}\,.
\end{split}
\end{equation}
Recalling that $D = 2y(m+n-1/y^2)$, the equation
\begin{equation}
	\der_y \ln T = D
\end{equation}
implies that
\begin{equation}
	t_1(\theta) = 0\,.
\end{equation}
Moreover we note that
\begin{equation}
	V_\phi \sim \frac{p_1 \cos^2\theta}{R^2} + \frac{V_2(\theta)}{R^4}\qquad V_r = 0\,.
\end{equation}
With a suitable coordinate transformation
\begin{equation}\label{vth}
\left \{ \begin{array}{l}
R = \sqrt{p_1} \tilde R + \frac{h_1(\tilde \theta)}{\tilde R}\\
\theta=\tilde\theta + \frac{g_1(\tilde\theta)}{\tilde R^2}\\
\phi = \tilde\phi+t
\end{array}
\right.
\end{equation}
it should be possible to bring the metric to the following form:
\begin{multline}\label{expectmetr}
		\dd s^2= \Omega(\tilde R , \tilde \theta)L^2 \left[- (1 + \tilde R^2 - \frac {\tilde R_0^2} {\tilde R^2})\dd t^2 + \frac {\dd \tilde R^2}{\tilde R^2} (1-\frac 1{\tilde R^2}) + \tilde R^2 \dd \tilde \Omega_3^2 \right]+\\
+L^2\bigg[ g_{\tilde{\theta}\tilde{\theta}}\dd \tilde \theta^2 + g_{\tilde{\phi}\tilde{\phi}}\cos^2\tilde\theta \left( \dd \tilde\phi + \frac {J} {\tilde R^2}\dd t\right)^2 +\\+  g_{\omega \omega}\sin^2\tilde\theta\dd \omega^2 + 
g_{\phi \phi}\sin^2\tilde\theta\bigg(\cos^2\omega (\dd \phi_1 - \frac {Q} {\tilde R^2}\dd t)^2  + \sin^2 \omega (\dd \phi_2 -  \frac {Q} {\tilde R^2}\dd t)^2   \bigg)\bigg]
\end{multline}
up to subleading corrections. To the leading order the metric components  $g_{\tilde{\theta}\tilde{\theta}}=g_{\tilde{\phi}\tilde{\phi}}=g_{\omega \omega}=g_{\phi \phi}=1$ and reproduce $S^5$. The constants $J$ and $Q$ are proportional to the total flux of the $U(1)$ gauge fields arising from the KK reduction of the supergravity over $S^5$. In particular $Q$ is the total charge of the solutions under both the gauge field associated with coordinate transformations generated by $\lambda(\xi) \der_{\phi_1}$ and $\mu(\xi) \der_{\phi_2}$ (being $\xi$ coordinates in the $AdS_5$ factor); these are dual respectively to the $J_1= L_{1,2}$ and $J_2=L_{3,4}$ $R$-symmetry generators. For this reason the expected BPS relation is
\begin{equation}
	M = \frac{\pi L^2}{4 G_5}(|J|+2| Q|)\,.
\end{equation}

The conformal factor $\Omega(\tilde R,\tilde \theta)$ satisfies \mbox{$\Omega(\tilde R=\infty,\tilde \theta)=1$} and contains terms up to order $\tilde R^{-4}$. The mass of the excitations over the $AdS_5$ vacuum is given by
\begin{equation}
	M = \frac{3 \pi L^2}{8 G_5} \tilde R_0
\end{equation}
where $G_5$ is the five-dimensional Newton constant\footnote{This approach follows the one in $\cite{bubbling}$. A more precise and detailed approach can be taken following e.g. the work in \cite{Skenderis:2006uy}}.
We recall now the expression for the metric:
\begin{multline}\label{ourmetr}
	\dd s^2 = - h^{-2} (\dd t^2 + V_\phi \dd\phi)^2 + h^2 \frac{\rho_1^2}{\rho_3^2}( T^2 \delta_{ij} \dd x^i \dd x^j + \dd y^2) + \\
+ \tilde \rho^2 \dd \Omega_3^2 + \rho_1^2 \big[\big(\sigma^{\hat 1}\big)^2+\big(\sigma^{\hat 2}\big)^2\big]+ \rho_3^2 (\sigma^{\hat 3}-A_t\dd t - A_\phi \dd \phi)^2=\\
= g_{tt} \dd t^2 + g_{RR} \dd R^2 + \tilde \rho^2 \dd \tilde \Omega_3^2 + 2 g_{\theta R} \dd \theta \dd R+\\ 
 + g_{t\tilde \phi}\dd t \dd \tilde \phi + g_{t\hat 3} \dd t \sigma^{\hat 3}+\\
+ g_{\theta\theta}\dd \theta^2 + g_{\tilde\phi\tilde\phi}\dd \tilde \phi^2+ g_{\tilde \phi \hat 3} \dd \tilde\phi \,\sigma^{\hat 3}+ \rho_1^2 \big[ \big(\sigma^{\hat 1}\big)^2+\big(\sigma^{\hat 2}\big)^2\big]+ \rho_3^2 \big(\sigma^{\hat 3}\big)^2
\end{multline}
with
\begin{equation}
\begin{split}
	&g_{tt} = - h^{-2} (1+V_\phi)^2 + h^2  \frac {\rho_1^2}{\rho_3^2}r^2 T^2 + \rho_3^2(A_\phi+A_t)^2\\
	&g_{RR} = h^2 \frac{\rho_1^2}{\rho_3^2}(\sin^2\theta+T^2 \cos^2 \theta)\\
	&g_{\theta R} = h^2 \frac{\rho_1^2}{\rho_3^2} R \sin\theta\cos\theta (1-T^2)\\
	&g_{t\tilde\phi} = - h^{-2} (1+V_\phi)V_{\phi}+h^2 \frac{\rho_1^2}{\rho_3^2}r^2 T^2 +\rho_3^2 (A_t+A_\phi)A_\phi\\
	&g_{t \hat 3} = - \rho_3^2 (A_t+A_\phi)\\
	&g_{\theta\theta} = h^2 \frac{\rho_1^2}{\rho_3^2} R^2 (\cos^2 \theta + T^2 \sin^2\theta)\\
	&g_{\tilde \phi \tilde\phi} = - h^{-2} V_{\phi}^2 +h^2 \frac{\rho_1^2}{\rho_3^2}r^2 T^2+\rho_3^2 A_\phi^2\\
	&g_{\tilde \phi \hat 3} = \rho_3^2 A_\phi
\end{split}
\end{equation}

We can now derive the $Q$ charge of our solutions. Using the definition \eqref{LIsigmahat} and the coordinate transformation \eqref{coordhat} we get
\begin{equation}
	Q = - \frac{ g_{t\hat 3}}{g_{\hat 3 \hat 3}}\tilde R^2 =  (A_t+A_\phi)\,.
\end{equation}
We note that $A_t = (n-p)/p =O( 1/R^2)$ and $A_\phi = A_t V_\phi+ \frac 12 r \der_r \ln T =O(1/R^4)$ and thus the leading behaviour of the r.h.s. is determined by $A_t$ and we have
\begin{equation}
	Q = \frac{n_2(\theta)-p_2(\theta)}{p_1^2}\,.
\end{equation}
Using these relations we can solve the equations \eqref{equations} up to 
second order in $\frac{1}{R^2}$ and demanding that the solutions are regular we find
\begin{equation}
\left\{
\begin{aligned}
&p_2(\theta) = d (3\cos^2\theta - 1)\\
&n_2(\theta) = p_2(\theta) + p_1^2 Q \\
&m_2(\theta) = -p_2(\theta) - 2 p_1^2 Q\\
&V_2(\theta)= \frac12 \cos^2\theta \big[(Q p_1^2  - d + 3 d \cos (2 \theta)\big]
\end{aligned}
\right.
\end{equation}
where $d$ is a generic real integration constant.
The $J$ charge is given by
\begin{equation}
	J= \frac{g_{t\tilde\phi}}{g_{\tilde\phi\tilde\phi}}\,\tilde R^2= \frac{d}{p_1^2}-1-Q.
\end{equation}
The conserved charges $Q$ and $J$ can be also obtained by evaluating Komar integrals 
associated with the Killing vectors $\hat{\Sigma}_3$ (the dual vector field to $\hat\sigma_3$) and $\frac{\partial}{\partial \phi}$
respectively.

We will now solve the equations to the next order and find the transformation \eqref{vth} that brings the metric to the form \eqref{expectmetr} enabling us to check that the BPS mass formula
\begin{equation}
	M = \frac{\pi L^2}{4 G_5}(|J|+2| Q|)
\end{equation}
is satisfied.

We have
\begin{equation}
	g_{\tilde \theta \tilde R} = [h_1'(\tilde\theta) -2 \sqrt{p_1} g_1(\tilde\theta)] \frac{1}{\tilde R^3}
\end{equation}
which fixes
\begin{equation}
	g_1 (\tilde\theta)= \frac{h_1'(\tilde\theta)}{2 \sqrt{p_1}}\, .
\end{equation}
We are not really interested in the conformal factor $\Omega(\tilde R,\tilde \theta)$ and we thus proceed to the calculation of the ratio
\begin{equation}
	\frac{g_{\tilde R \tilde R}}{\tilde \rho^2} =  \frac{1}{\tilde R^4} + \frac{d(3\cos^2\tilde\theta-1)-6 p_1^{3/2} h_1(\tilde\theta)}{p_1^2}
\end{equation}
 which should satisfy the equation
\begin{equation}
	\frac{g_{\tilde R \tilde R}}{\tilde \rho^2} = \frac{1}{\tilde R^4}-\frac{1}{\tilde R^6}\,.
\end{equation}
This requirement gives immediately, 
\begin{equation}
	 h_1(\tilde\theta) = \frac{p_1^2 +d(3\cos^2\tilde\theta-1) }{6 p_1^{3/2}}\,.
\end{equation}
Using this relation we obtain
\begin{equation}
	\frac{g_{tt}}{\tilde\rho^2} = -1 -\frac{1}{\tilde R^2}+\frac{2}{3}\left( \frac{d}{p_1^2}-1-3Q\right) \frac{1}{\tilde R^4}
\end{equation}
which gives
\begin{equation}
	\tilde R_0 = \frac23 (J -2 Q)
\end{equation}
and thus
\begin{equation}
	M = \frac{3 \pi L^2}{8 G_5} \tilde R_0 = \frac {\pi L^2} {4 G_5} (J-2 Q)\,.
\end{equation}
This should be compared to
\begin{equation}
	M = \frac{\pi L^2 }{4 G_5}(|J|+2| Q|)\,,
\end{equation}
which apparently requires that $J>0$ and $Q<0$. Up to now, $J$ and $Q$ have appeared in the solution to the differential equations as constants of integration. As such, they can take any real value. Constraints on their possible values should come from a global analysis of the solutions
\footnote{As in the LLM case, the sign
of $J$ is correlated with the relative chirality of the Killing spinor with respect to the two $SO(4)$'s. From the gauge theory side, as follows from the discussion at the end of Section 2, the sign of $Q$ is correlated with the $U(1)_R$ charge of the Killing spinor. As it emerges from the detailed analysis of Appendix B, this charge is captured by the eigenvalue
$s$ with respect to a Pauli matrix $\sigma_{\hat 3}$. In our analysis we have set for definiteness $s=+1$. Had we chosen $s=-1$, $Q$ would have been positive.}. Indeed given the leading behaviour at large $R$, these subleading corrections should be completely determined by the boundary conditions at $y=0$. Unfortunately we are not able to express these charges in terms of the data at $y=0$ plane which could have allowed us to establish the above bounds on $J$ and $Q$. As a matter of comparison, in the LLM construction only the $J$ charge is present and its value is determined by a set of integrals performed on the $y=0$ plane. In that case, the bound $J>0$ is trivially imposed by the specific type of boundary conditions at $y=0$.
\section{Conclusions and perspectives}
In this paper we have extended to the 1/8 BPS case the construction of \cite{bubbling}.
 Due to the reduced amount of symmetry of our background
the expressions we find  turn out to be rather more complex; 
in particular the differential equations which determine the background are highly non linear. 
We performed an asymptotic analysis for large values of $R$ and were able to show 
that solutions with the desired asymptotics and regularity exist in this limit.
Of course, a satisfactory understanding of the boundary conditions at $y=0$ which lead to non-singular solutions is necessary in order
to connect the geometry of the supergravity solutions to the phase space of the
quantum mechanical system arising from the dual gauge theory on $\RR \times S^3$.
In particular it would be very interesting to understand the relationship between our construction 
and the work of \cite{Berenstein:2005aa,Mikhailov:2000ya,Biswas:2006tj,Mandal:2006tk}. 
Once the space of solutions is understood from the supergravity point of view 
one could proceed to its quantisation by a procedure like that
presented in \cite{Grant:2005qc,Maoz:2005nk}.

Our solutions have a non empty intersection with the solutions described in 
\cite{Donos:2006ms,Donos:2006iy} and in \cite{Chong:2004ce}. It would be interesting 
to find the exact dictionary between different descriptions of the same solutions 
in order to better clarify the role of the boundary conditions at $y=0$ and to try to 
recast the differential equations in a more tractable fashion. 
$1/4$ BPS solutions can be obtained from the general setting that we have presented by 
imposing some additional constraints on the four scalar functions \cite{future}. 
Some of the so-called superstar geometries in \cite{Myers:2001aq} are also contained in 
our description. These solutions are known to have singularities and it is possible to 
identify the boundary conditions at $y=0$ that are responsible for them. With a more 
detailed understanding of boundary conditions which give rise to 
non-singular solutions, and their relation to the CFT, one may better understand the 
resolution of the singularities in a manner similar to that of 
\cite{Caldarelli:2004mz,Milanesi:2005tp,Balasubramanian:2005mg}. 
Finally different types of boundary conditions at large $R$ can be studied. 
Indeed one can find solutions with asymptotics of the 
form $AdS_5 \times Y^{p,q}$: such geometries correspond to $1/2$ BPS operators in the 
$\mathcal{N}=1$ superconformal quiver gauge theories\cite{future2}.

\subsubsection*{Acknowledgements}
 This work
was supported in part by the European Commission under the RTN contract MRTN-CT-2004-503369.
M.O'L. would like to thank the ICTP for hospitality during the course of this work.
The work of G.M. is supported in part by the EC under the contract MRTN-CT-2004-005104.

\appendix

\section{Conventions}

We set up our conventions for the wedge product of 1-forms
\begin{equation}
	\alpha_1\wedge\cdots \wedge \alpha_n = \frac{1}{n!} \sum_{i}\sigma(i)\alpha_{i(1)}\otimes\cdots \otimes \alpha_{i(n)}
\end{equation}
where the sum is over the $n!$ permutations $i$ and $\sigma(i)$ is the parity of the permutation.

An $n$-form $\alpha$ in a $d$ dimensional space $(\alpha \in \Lambda_n)$ is given by
\begin{equation}
	\alpha = \bar\alpha_{\mu_1 \cdots \mu_n} \dd x^{\mu_1}\wedge\cdots \wedge \dd x^{\mu_n}=\frac{1}{n!} \alpha_{\mu_1 \cdots \mu_n} \dd x^{\mu_1}\wedge\cdots \wedge \dd x^{\mu_n}
\end{equation}
with $\alpha_{\mu_1 \cdots \mu_n} $ the complete antisymmetrization 
of $\bar\alpha_{\mu_1 \cdots \mu_n}$.

When a metric is present we can introduce the Hodge dual
\begin{equation}
	\star : \Lambda_n \tend \Lambda_{d-n}
\end{equation}
Given a $d$-bein of the metric $\{ e^m\}_{m=1,\cdots d}$,
\begin{equation}
	\star e^{m_1}\wedge\cdots\wedge e^{m_n} = \frac{1}{(d-n)!}\epsilon^{m_1,\dots m_n,m_{n+1},\cdots, m_d} e_{m_{n+1}}\wedge\cdots\wedge e_{m_d}
\end{equation}
where indices are lowered with the tangent space metric. From this definition it follows that
\begin{multline}
	\star \dd x^{\mu_1}\wedge \cdots \wedge \dd x^{\mu_n} = \star g^{\mu_1 \mu'_1}\cdots g^{\mu_n\mu'_n} e_{m_1\mu'_1}\cdots e_{m_n\mu'_n} e^{m_1}\wedge\cdots\wedge e^{m_n} = \\ =  g^{\mu_1 \mu'_1}\cdots g^{\mu_n\mu'_n} e_{m_1\mu'_1}\cdots e_{m_n\mu'_n}\frac{1}{(d-n)!}\epsilon^{m_1,\cdots m_n,m_{n+1},\cdots, m_d} e_{m_{n+1}}\wedge\cdots\wedge e_{m_d}=
\\= \frac{1}{(d-n)!}\sqrt{g}g^{\mu_1 \mu'_1}\cdots g^{\mu_n\mu'_n}\epsilon_{\mu'_1,\cdots \mu'_n,\mu'_{n+1},\cdots, \mu'_d}\dd x^{\mu'_{n+1}}\wedge\cdots\wedge \dd x^{\mu'_d}\,.
\end{multline}
The exterior derivative of a 1-form is defined by
\begin{equation}
	\beta = \dd \alpha = \der_\mu \alpha_\nu \dd x^\mu \wedge \dd x^\nu = \frac 12 (\der_\mu\alpha_\nu -\der_\nu\alpha_\mu ) \dd x^\mu \wedge \dd x^\nu 
\end{equation}
or in terms of components $\beta_{\mu\nu} = \der_\mu\alpha_\nu -\der_\nu\alpha_\mu$.
The generalization to any $n$-form is given by
\begin{multline}
	\beta = \dd \alpha = \frac{1}{n!} \der_\mu \alpha_{\nu_1\cdots\nu_n}\dd x^\mu\wedge\dd x^{\nu_1}\wedge\cdots\wedge\dd x^{\nu_n} =\\= \frac{1}{(n+1)!}\der_{[\mu}\alpha_{\nu_1\cdots\nu_n]}\dd x^\mu\wedge\dd x^{\nu_1}\wedge\cdots\wedge\dd x^{\nu_n}=\frac{1}{(n+1)!}\beta_{\nu_1\cdots\nu_n}\dd x^{\nu_1}\wedge\cdots\wedge\dd x^{\nu_n}
\end{multline}
where now $\beta_{\mu\nu_1\cdots\nu_n} = \der_{[\mu}\alpha_{\nu_1\cdots\nu_n]}$ and 
square brackets indicate antisymmetrization without normalization.

The torsionless spin connection 1-form is defined by the structure equation:
\begin{equation}
	\dd e^a +\omega^a_{\phantom a b}\wedge e^b = 0\,.
\end{equation}
Requiring metricity of the connection
\begin{equation}
	\omega_{ab} = - \omega_{ba}
\end{equation}
allows us to explicitly express $\omega_{ab}$ in terms of the $d$-bein  ($E_a$ are the inverse $d$-bein vector fields, defined by $e^a \cdot E_b = \delta^a_{ \phantom b a}$),
\begin{multline}
	\omega_{ab} = -\dd e_a \cdot E_b + \dd e_b \cdot E_a + \frac 12 \left(e^c \cdot \left[ E_a, E_b\right]\right) e_c= \\ =
\Big[-\frac12 \left(\der_\mu e_{a \nu} - \der_\nu e_{a \mu}\right) E_{\phantom\nu b}^{\nu} + \frac12 \left(\der_\mu e_{b \nu} - \der_\nu e_{b \mu}\right) E_{\phantom \nu a}^{\nu}+\\+\frac 12 e^c_{\phantom c \rho} \left(E_{\phantom \nu a}^{\nu} \der_\nu E_{\phantom \rho b}^{\rho}-E_{\phantom \nu b}^{\nu}\der_\nu E_{\phantom \rho a}^{\rho}\right)e_{c\mu}\Big]\dd x^\mu = \\
 = \Big[-\frac12 \left(\der_\mu e_{a \nu} - \der_\nu e_{a \mu}\right) E_{\phantom\nu b}^{\nu} + \frac12 \left(\der_\mu e_{b \nu} - \der_\nu e_{b \mu}\right) E_{\phantom \nu a}^{\nu}+\\-\frac12 E_{\phantom \nu a}^{\nu} \left(\der_\nu e^c_{\phantom c \rho}-\der_\rho e^c_{\phantom c \nu}\right)E_{\phantom \rho b}^{\rho} e_{c\mu}\Big]\dd x^\mu = \\
= -\dd e_a \cdot E_b + \dd e_b \cdot E_a - \left(E_a \cdot \dd e^c \cdot E_b\right) e_c
\end{multline}
where in going from the second to the third line we have used
\begin{equation}
	0 = \der_\mu \eta_{ab} = \der_\mu \left(e_{a \nu} E^{\nu}_{\phantom \nu b}\right) = \left(\der_\mu  e_{a \nu} \right)  E^{\nu}_{\phantom \nu b}+ e_{a \nu}\left(\der_\mu E^{\nu}_{\phantom \nu b}\right)\,.
\end{equation}
This is an explicit realization of the identity
\begin{equation}\label{exterior1forms}
	V\cdot \dd \alpha  \cdot W = \frac{1}{2} \dd (\alpha\cdot W) \cdot V -\frac12 \dd (\alpha\cdot V) \cdot W - \frac12 \alpha \cdot [V,W]\,,
\end{equation}
which holds for any one form $\alpha$ and any pair of vector fields $V,W$.

The covariant derivative of a spinor is given by
\begin{equation}
	\nabla_\mu \psi = \der_\mu \psi + \frac 14 \omega_{ab \mu} \Gamma^a\Gamma^b \psi\,.
\end{equation}

\subsubsection*{Group manifolds}
Consider a Lie algebra of vector fields on a $d$-dimensional group manifold. 
It is a $d$ dimensional vector space of vector fields satisfying
\begin{equation}
	[E_a,E_b] = f_{ab}^{\phantom{ab} c} E_c\,.
\end{equation}
The exterior derivative of the dual one forms is given by
\begin{equation}
	\dd e^c = \frac12 \alpha_{ab}^{\phantom{ab}c} e^a\wedge e^b
\end{equation}
These are the Maurer Cartan 1-forms. Indeed, we have 	
\begin{equation}
	E_a\cdot \dd e^c \cdot E_b = \frac 12 \alpha_{ab}^{\phantom{ab}c}
\end{equation}
and according to \eqref{exterior1forms}
\begin{equation}
	E_a\cdot \dd e^c \cdot E_b = -\frac 12 e^c \cdot\left[E_a,E_b\right]
\end{equation}
which give
\begin{equation}
	\alpha_{ab}^{\phantom{ab}c} = - f_{ab}^{\phantom{ab}c}\,.
\end{equation}
The Lie derivative of a 1-form is defined by
\begin{equation}
	(\mathcal L_{J} \omega)\cdot K = \der_K (\omega\cdot K) - \omega \cdot [J,K]
\end{equation}
and thus
\begin{equation}
	\mathcal L_{E_a} e^c = - f_{ab}^{\phantom{ab}c} e^b 
\end{equation}
Taking these $e^a$ as the $d$-bein, the spin connection on the group manifold is given by
\begin{equation}
	\omega_{abc} = \frac12 (-\alpha_{cba}+\alpha_{cab}+f_{abc}) = \frac12 (f_{cba}-f_{cab}+f_{abc})\,.
\end{equation}

\section{Reduction of the Killing spinor equations}
In this Appendix we present the step by step derivation of the results presented in Section 3.
\subsection{Metric and 5-form Ansatz}
The most generic Ansatz for our solutions is given by
\begin{equation}
	\dd s^2 = g_{\mu\nu} \dd x^\mu \dd x^\nu + \rho_1^2 \big[(\sigma^{\hat 1})^2 + (\sigma^{\hat 2})^2\big] +\rho_3^2(\sigma^{\hat 3} - A_\mu \dd x^\mu)^2 + \tilde \rho^2 \dd \tilde \Omega_3^2\,.
\end{equation}
The space is thus made up of a fibration over a four dimensional manifold of a squashed 3-sphere (on which the $SU(2)$ left-invariant 1-forms $\sigma^{\hat a}$ are defined) and a round 3-sphere (on which the $SU(2)$ left-invariant 1-forms $\sigma^{\tilde a}$ are defined).\\
The left invariant 1-forms are given by
\begin{equation}
\begin{array}{ll}
\sigma^{\hat 1}= -\frac 12 (\cos \hat\psi\, d\hat\theta + \sin \hat\psi\, \sin\hat \theta\, d\hat \phi) & \sigma^{\tilde 1}= -\frac12(\cos \tilde\psi\, d\tilde\theta + \sin \tilde\psi\, \sin \tilde\theta\, d\tilde\phi)\\
\sigma^{\hat 2}= -\frac12(-\sin\hat \psi\, d\hat\theta + \cos\hat \psi\, \sin\hat \theta\, d\hat \phi )& \sigma^{\tilde 2}= -\frac12(-\sin \tilde\psi\, d\tilde\theta + \cos \tilde\psi\, \sin\tilde \theta\, d\tilde\phi) \\
\sigma^{\hat 3}= -\frac12(d\hat\psi + \cos \hat\theta\, d \hat\phi) & \sigma^{\tilde 3}= -\frac12(d\tilde\psi + \cos \tilde\theta\, d \tilde\phi) 
\end{array}
\end{equation}
and satisfy the relations
\begin{equation}
\begin{split}
	\dd \sigma^{\hat i } = \epsilon_{\hat i \hat j \hat k} \sigma^{\hat j }\wedge \sigma^{\hat k}\\
\dd \sigma^{\tilde i} = \epsilon_{\tilde i \tilde j \tilde k} \sigma^{\tilde j}\wedge \sigma^{\tilde k}\,.
\end{split}
\end{equation}
With this normalization the metric on the unit radius round three sphere is given by
\begin{equation}
	d \Omega_3^2 = (\sigma^1)^2+(\sigma^2)^2+(\sigma^3)^2
\end{equation}
with $\sigma^a$ being either $\sigma^{\hat a}$ or $\sigma^{\tilde a}$.\\
We choose our ``d-bein'' to be
\begin{align}
e^m =& \veps{m}{\mu}\dd x^\mu\\
e^{\hat a} =& \begin{cases}
	\rho_1 \sigma^{\hat a} & a=1,2\\
	\rho_3(\sigma^{\hat 3} -A_\mu\dd x^\mu) & a=3
      \end{cases}\\
e^{\tilde a} =& \tilde \rho \sigma^{\tilde{a}}
\end{align}
The only non zero Ramond-Ramond field strength is the five form $F_{(5)}$ and the dilaton is assumed to be constant. The most generic Ansatz for the five form which is invariant under the given symmetries is
\begin{multline}
	F_{(5)} =  2 \left( \tilde G_{mn} e^m\wedge e^n + \tilde V_m e^m\wedge e^{\hat 3} + \tilde g e^{\hat 1}\wedge e^{\hat 2} \right)\wedge \tilde\rho^3\dd \tilde \Omega_3+\\
	2\left(- G_{pq}e^p\wedge e^q \wedge e^{\hat 1}\wedge e^{\hat 2}\wedge  e^{\hat 3} +\star_4 \tilde V\wedge  e^{\hat 1}\wedge e^{\hat 2}- \star_4 \tilde g \wedge e^{\hat 3}\right) \,,
\end{multline}
where
\begin{gather}
	G_{mn} = \frac12 \epsilon_{mnpq} \tilde G^{mn}\\
	\star_4 \tilde V = \frac{1}{3!}\epsilon_{mnpq}\tilde V^m e^n\wedge e^p\wedge e^q\\
	\star_4 \tilde g  = \tilde g e^0\wedge e^1\wedge e^2\wedge e^3\,.
\end{gather}
The Bianchi identity $\dd F_{(5)} = 0$ gives rise to the set of equations,
\begin{gather}
	\dd \big(\tilde G \tilde \rho^3 - \tilde V \wedge A \rho_3\tilde\rho^3\big) = 0\\
	\tilde V = \frac 12 \frac{1}{\rho_3 \tilde\rho^3}\dd (\tilde{g} \rho_1^2 \tilde\rho^3)\\
	\dd \big( G \rho_1^2 \rho_3\big) = 0\\
	\dd \big( G \rho_1^2 \rho_3 \wedge A + \star_4 \tilde V \big)-2\star_4 \tilde g = 0\,.
\end{gather}
\subsection{Spin Connection and Covariant Derivative.}
The inverse d-bein is
\begin{align}
E_m =& \Xi^{\mu}_{\phantom \mu m}\der_\mu+A_m\Sigmahat{i}{3}\der_{\hat i}\\
E_{\hat a} =& \frac{1}{\rho_a}\Sigmahat{i}{a}\der_{\hat i}\\
E_{\tilde a} =& \frac{1}{\tilde\rho}\Sigmatilde{i}{a}\der_{\tilde i}\,,
\end{align}
where  $\Xi_m$ is the inverse vierbein of $\eps^m$ and $\Sigma_{\hat a, \tilde a}$ is the 
inverse of $\sigma^{\hat a, \tilde a}$.
We will denote ten-dimensional tangent space indices by $A,B,C...$.
The spin connection is given by
\begin{equation}
	\omega_{AB} = -\dd e_A\cdot E_B + \dd e_B \cdot E_A + \frac 12 \left(e^C \cdot \left[ E_A, E_B\right]\right) e_C\,.
\end{equation}
Using the explicit expressions for $E_m$ we have
\begin{equation}
[E_m,E_n] = [\Xi_m,\Xi_n] 
+ \Sigma_{\hat 3} \left(\Xi_m (A \cdot \Xi_n) - \Xi_n (A\cdot \Xi_m)\right) \,.
\end{equation}
We can thus write, using the relation \eqref{exterior1forms}
\begin{multline}
	\omega_{mn} = \tilde \omega_{mn} +e^{\hat 3} \rho_3\frac12 \left(-A\cdot [\Xi_m,\Xi_n]+\Xi_m (A \cdot \Xi_n)- \Xi_n (A\cdot \Xi_m)\right) = \\
	=  \tilde \omega_{mn} + e^3 \rho_3 \Xi_m \cdot \dd A \cdot \Xi_n\,.
\end{multline}
In order to get the other components of the spin connection we will need the explicit form of the exterior derivative of $e^{\hat a} = \rho_{\hat a} \sigma^{\hat a} - \rho_3 \delta^{\hat a}_{\hat 3} A_m e^m$ and of $e^{\tilde a} = \tilde \rho \sigma^{\tilde a}$
\begin{gather}
	\dd e^{\hat a} = \dd \rho_{\hat a} \wedge \sigma^{\hat a} + \rho_{\hat a} \dd\sigma^{\hat a} - \rho_3 \delta^{\hat a}_{\hat 3} \dd A - \delta^{\hat a}_{\hat 3}  \dd \rho_3 \wedge A_m e^m\\
\dd e^{\tilde a} = \dd \rho_{\tilde a}\wedge \sigma^{\tilde a} + \rho_{\tilde a} \dd\sigma^{\tilde a} 
\end{gather}
By definition
\begin{equation}
\begin{split}
	\dd \sigma^{\hat i } = \epsilon_{\hat i \hat j \hat k} \sigma^{\hat j }\wedge \sigma^{\hat k}\\
\dd \sigma^{\tilde i} = \epsilon_{\tilde i \tilde j \tilde k} \sigma^{\tilde j}\wedge \sigma^{\tilde k}
\end{split}
\end{equation}
and thus
\begin{equation}
\begin{split}
	[\Sigma_{\hat a},\Sigma_{\hat b}] = - 2 \epsilon_{\hat c \hat a \hat b}  \Sigma_{\hat c}\\
[\Sigma_{\tilde a},\Sigma_{\tilde b}] = - 2 \epsilon_{\tilde c \tilde a \tilde b}  \Sigma_{\tilde c}\,,
\end{split}
\end{equation}
so that
\begin{equation}
[E_{\hat a},E_m] = \frac{1}{\rho_a^2}\der_m \rho_a \Sigma_{\hat a}+\frac{1}{\rho_a} A_m [\Sigma_{\hat a},\Sigma_{\hat 3}] =\frac{1}{\rho_a^2}\der_m \rho_a \Sigma_{\hat a}- \frac{2}{\rho_a} A_m   \epsilon_{\hat c \hat a \hat 3}  \Sigma_{\hat c}\,.
\end{equation}
In the end
\begin{multline}
\omega_{\hat am} = -\dd e_{\hat a} E_m + \frac 12 e^P [E_{\hat a},E_m] e_P = \\=
\der_m\rho_a \sigma^{\hat a}+\delta^{\hat a}_{\hat 3} e^p (\rho_3\frac12  F_{pm}-A_p\der_m\rho_3)
\end{multline}
\begin{equation}
	\omega_{\tilde am} = -\dd e_{\tilde a} E_m + \frac 12 e^P [E_{\tilde a},E_m] e_P = \der_m\tilde \rho\, \sigma^{\tilde a}
\end{equation}
and
\begin{multline}
\omega_{\hat a\hat b} = -\dd e_{\hat a} E_{\hat b} + \dd e_{\hat b} E_{\hat a} + \frac12 e^M \cdot [E_{\hat a},E_{\hat b}] e_M = \\=
\epsilon_{\hat a\hat b\hat c} \left(\frac{\rho_a^2+\rho_b^2-\rho_c^2}{\rho_a\rho_b}\right)\sigma^{\hat c}+\epsilon_{\hat a\hat b\hat 3}\frac{\rho_3^2}{\rho_1^2}A
\end{multline}
\begin{equation}
	\omega_{\tilde a\tilde b} = -\dd e_{\tilde a} E_{\tilde b} + \dd e_{\tilde b} E_{\tilde a} + \frac12 e^M \cdot [E_{\tilde a},E_{\tilde b}] e_M=\epsilon_{\tilde a\tilde b\tilde c}\sigma^{\tilde c}\,.
\end{equation}
The spin connection part of the covariant derivative acting on spinors as presented in Appendix A is
\begin{equation}
\begin{split}
&\frac14 \omega_{MN}\Gamma^M\Gamma^N =\\ &\dd x^\mu \Bigg[\frac14 \tilde \omega_{mn} \Gamma^m\Gamma^n -\frac14\rho_3 F_{\mu m} \Gamma^m\Gamma^{\hat 3}+\\&\qquad\qquad-A_\mu \left(-\frac12\frac{\rho_3^2}{\rho_1^2}\Gamma^{\hat 1}\Gamma^{\hat 2}-\frac12\der_m \rho_3\Gamma^m\Gamma^{\hat 3}+ \frac 18 \rho_3^2  F_{mn} \Gamma^m\Gamma^n \right)\Bigg]+\\
&\sum_{a=1,2}\sigma^{\hat a}\left(\frac12 \frac{\rho_3}{\rho_1}\epsilon_{\hat a \hat b \hat 3}\Gamma^{\hat b}\Gamma^{\hat 3}-\frac12\der_m\rho_1 \Gamma^m\Gamma^{\hat a}\right)+\\&\quad+
\sigma^{\hat 3} \left(\frac12 \left(2-\frac{\rho_3^2}{\rho_1^2}\right)\Gamma^{\hat 1}\Gamma^{\hat 2}-\frac12\der_m\rho_3\Gamma^m\Gamma^{\hat 3}+\rho_3^2\frac18 F_{mn}\Gamma^m\Gamma^n\right)+\\&
\sum_{a=1,2,3}\sigma^{\tilde a} \left( \frac 14 \epsilon_{\tilde a \tilde b \tilde c}\Gamma^{\tilde b}\Gamma^{\tilde c}-\frac 12 \der_m\tilde \rho \Gamma^m\Gamma^{\tilde a}\right)
\end{split}
\end{equation}
\subsection{Killing spinor}
\subsubsection*{Conventions and Ansatz}
We choose the following ten dimensional gamma matrices
\begin{equation}\begin{split}
	&\Gamma_m = \gamma_m\otimes 1 \otimes 1 \otimes 1\quad \Gamma^{\hat a} = 1\otimes\hat \sigma_1 \otimes \sigma_{\hat a}\otimes 1\quad \Gamma^{\tilde a} = 1\otimes\hat \sigma_2 \otimes 1 \otimes \sigma_{\tilde a}
\end{split}
\end{equation}
The two 32 component Majorana-Weyl spinor supersymmetry parameters of the IIB theory can be 
grouped into a single complex Weyl spinor $\psi$ obeying the chirality constraint
\begin{gather}\label{chiralproj}
	\Gamma_{11} \psi = \psi \\
 \Gamma_{11} = \prod_m \Gamma_m \prod_{\hat a} \Gamma_{\hat a} \prod_{\tilde a}\Gamma_{\tilde a} = \gamma_5 \hat \sigma_3 \qquad \gamma_5= -\ii \gamma_0\gamma_1\gamma_2\gamma_3\,.
\end{gather}

The supersymmetry variation of the gravitino $\chi_M$ is given by
\begin{equation}
	\delta \chi_M = \nabla_M \psi + \frac{\ii}{480}F_{M_1 M_2 M_3 M_4 M_5} \Gamma^{M_1 M_2 M_3 M_4 M_5}\Gamma_M \psi\,.
\end{equation}
In order to have a supersymmetric background we need to impose that this variation is 
zero giving rise to the Killing spinor equation on $\psi$,
\begin{equation}
	\nabla_M \psi + \frac{\ii}{480}F_{M_1 M_2 M_3 M_4 M_5} \Gamma^{M_1 M_2 M_3 M_4 M_5}\Gamma_M \psi= 0\,.
\end{equation}

As a consequence of our symmetry assumptions we look for a $\psi$ of the form
\begin{equation}
	\psi = \eps_{(b)}  \otimes \hat \chi \otimes \tilde \chi_{(b)}\,.
\end{equation}
Where $\psi$ is an 8 component complex spinor a and $\hat \chi, \tilde\chi_b$ are 2 components complex spinors defined on the two 3-spheres satisfying
\begin{gather}
	\Sigma_{\hat a} \hat\chi = 0 \qquad \sigma_{\hat 3} \hat\chi = s \chi\\
	\nabla'_{\tilde a}\tilde\chi = b \frac{\ii}{2}\sigma_{\tilde a} \tilde \chi_{(b)}
\end{gather}
where $\nabla'$ is the covariant derivative on the unit radius three sphere which has spin connection $\omega'_{abc} = \epsilon_{abc}$ and $s,b=\pm 1$. As we are going to show in the following, this choice means that $\hat \chi$ is a constant spinor and thus a singlet of the $SU(2)_L$ isometry of the squashed sphere, as required by our analysis of the gauge theory description of supersymmetries in Section \ref{gaugetheory}.
\subsubsection*{Isometries and Spinors}
On a unit radius round three sphere there exist two linearly independent solutions to the equation
\begin{equation}
	\nabla_{ a}\chi = \beta \frac{\ii}{2}\sigma_{a} \chi
\end{equation}
for each choice of $\beta=\pm 1$. The sign of $\beta$ is correlated with the chirality of the doublet of solutions under the $SO(4) = SU(2)\times SU(2)$ isometry group of $S^3$. This can be understood as follows.\\
Given a $d$-bein $e^a (y)$ and an isometry $I$ we choose a local orthogonal transformation $\Lambda$ such that
\begin{equation}
	\Lambda^{a}_{\phantom a b} \mathrm T I_{*} (e^b) = e^a 
\end{equation}
where $\mathrm T I_*$ is the pullback of one forms associated with $I$.
The $d$-bein is thus invariant under these transformations and it is possible to give meaning to the transformation properties of spinors under the isometries of the metric.\\
In our case, since $S^3\thickapprox SU(2)$, we can identify the points $y$ with elements of $SU(2)$. For the round 3-sphere $\tilde S^3$ the action of the isometry group $SU(2)_L\times SU(2)_R$ is 
given by left and right multiplication by generic elements of $SU(2)$. For the squashed three sphere the action of the isometry group $SU(2)_L\times U(1)_R$ is given by left multiplication by 
generic elements of $SU(2)$ and right multiplication with a $U(1)$ subgroup. \\
Let's focus on the left isometries $L_g$. They are defined by
\begin{equation}
	L_g (y) = g y\,.
\end{equation}
As our 3-bein is built out of left-invariant one forms $\sigma^a$, we have by definition
\begin{equation}
	\mathrm T L_{g\, *}( \sigma^a) = \sigma^a
\end{equation}
which implies that, for such transformations, $\Lambda^a_{\phantom a b} = \delta^a_{\phantom a b}$. The action $\mathrm S L_g$ on spinors of this isometry is thus very simple
\begin{equation}
	\mathrm S L_g \chi (g y) = \chi (y)\,.
\end{equation}
The action of left multiplications is clearly surjective and thus a spinor $\chi$ is invariant under this action if and only if it is a constant spinor. This means that our spinor $\hat \chi$ is a singlet under the $SU(2)_L$ isometry of the squashed 3-sphere, while the spinors $\tilde \chi_\pm$ transform in the $(0,\frac 12)$ for upper sign and $(\frac 12, 0)$ for the lower sign. For a discussion of spinors in squashed 3-spheres see \cite{Gibbons:1979kq,Hitchin}.
\subsubsection*{Equations and bilinears}
We turn now to the contribution of the Ramond-Ramond form to the gravitino variation. We define
\begin{equation}
M \equiv \frac{\ii}{480}F_{M_1 M_2 M_3 M_4 M_5} \Gamma^{M_1 M_2 M_3 M_4 M_5} .
\end{equation}
The chirality condition on $\psi$ and the self-duality of $F_{(5)}$ imply that
\begin{equation}
	M \Gamma_M \psi = -\bigg( \tilde G \sla + \tilde V \sla \gamma_5 \hat \sigma_1\sth  + \ii \tilde g \sth\bigg) \gamma_5\hat \sigma_2 \Gamma_M\psi\,.
\end{equation}
Due to the conditions on the spinor, $\hat \chi$ and $\tilde \chi_b$ factorise in each component of the gravitino variation equation which then becomes the following system of coupled differential and algebraic equations on $\eps$\footnote{For example the first equation is obtained as follows
\begin{multline}
	\left(\nabla_\mu+M \Gamma_\mu \right)\psi = \left(\tilde\nabla_\mu -\frac14\rho_3 F_{\mu\nu}\Xi^\nu_{\phantom \nu m}\Gamma^m\Gamma^{\hat 3}+A_\mu\left(\Sigma_{\hat 3}+\Gamma^{\hat 1}\Gamma^{\hat 2}\right)-
A_\mu \nabla_{\hat 3}+M \Gamma_\mu\right)\psi=\\
=\left(\tilde\nabla_\mu -\frac14\rho_3 F_{\mu\nu}\Xi^\nu_{\phantom \nu m}\Gamma^m\Gamma^{\hat 3}+A_\mu\Gamma^{\hat 1}\Gamma^{\hat 2}+M\left(\Gamma_\mu+A_\mu\rho_3\Gamma_{\hat 3}\right) \right)\psi= \\
=\left(\tilde\nabla_\mu -\frac14\rho_3 F_{\mu\nu}\Xi^\nu_{\phantom \nu m}\gamma^m\sigma^{\hat 3}+A_\mu\sigma_{\hat 3}+M\gamma_\mu\right)\psi
\end{multline}}
\begin{align}
	&\left[\tilde \nabla_\mu - \frac 14 F_{\mu\nu}\Xi^\nu_{\phantom \nu m} \gamma^m\gamma^5 \hat \sigma_1 s+ \ii A_\mu s - \bigg( \tilde G \sla + \tilde V \sla \gamma_5 \hat \sigma_1 s + \ii \tilde g s\bigg) \gamma_5\hat \sigma_2 \gamma_\mu \right]\eps=0\label{mu}\\
&\left[\frac \ii 2 \frac{\rho_3}{\rho_1}\gamma_5\hat\sigma_1+\frac12 \der\,\sla\rho_1+\rho_1\bigg( \tilde G \sla + \tilde V \sla \gamma_5 \hat \sigma_1 s  - \ii \tilde g s \bigg) \gamma_5\hat \sigma_2\right]\eps=0\label{a12}\\
&\left[\frac{\ii}{2}\left(2-\frac{\rho_3^2}{\rho_1^2}\right) \gamma_5\hat \sigma_1 +\frac12 \der\,\sla \rho_3+\frac18 \rho_3^2 F\sla\gamma_5\hat\sigma_1 s +\rho_3\bigg( \tilde G \sla - \tilde V \sla \gamma_5 \hat \sigma_1 s  + \ii \tilde g s \bigg) \gamma_5\hat \sigma_2 \right]\eps=0\label{a3}\\
&\left[\frac \ii 2 b\gamma_5\hat\sigma_2 + \frac12 \der\,\sla \tilde \rho -\tilde\rho \bigg( \tilde G \sla + \tilde V \sla \gamma_5 \hat \sigma_1 s  + \ii \tilde g s \bigg)\gamma_5\hat\sigma_2\right]\eps=0\label{at}. 
\end{align}
Note that the first equation is a  first order differential 4-vector equation for $\eps$ while the last three are  algebraic 4-scalar equations.\\
We now define a useful set of bilinears
\begin{equation}
\begin{split}
    &K_\mu=\bar\eps\gamma_\mu\eps\qquad
    L_\mu=\bar\eps\gamma_5\gamma_\mu\eps\qquad Y_{\mu\nu}=\bar\eps\gamma_{\mu\nu}\sigma_1\eps\\
    &f_1=\ii\bar\eps\sigma_1\eps\qquad f_2=\ii\bar\eps\sigma_2\eps\\
    &\bar\eps=\eps^\dag\gamma_0
  \end{split}
\end{equation}
The world indices $\mu,\nu$ of these bilinears are obtained by contraction of the tangent space indices with the vierbein $\eps^m_{\phantom m \mu}$. When raising and lowering $\mu$ indices we will always use the metric $\tilde g _{\mu\nu}$ unless otherwise is specified.
By Fierz rearrangements the following relations can be proved
\begin{equation}
		K^2 = - L^2 = -f_1^2-f_2^2 \equiv - h^{-2}\qquad L^\mu K_\mu = 0
\end{equation}

\subsection{Algebraic relations}
By multiplying the algebraic equations \eqref{a12},\eqref{a3},\eqref{at} with different combinations of gamma matrices and contracting with $\bar\eps$ one can obtain the following relations for the 
spinor bilinears:
\begin{gather}
	K^\mu \der_\mu\rho_1 = 0\\
	K^\mu \der_\mu\rho_3 = 0\\
	K^\mu \der_\mu\tilde\rho=0\\
	L_\mu = - \frac{\rho_1}{\rho_3} \frac{f_1}{\tilde\rho} \der_\mu (\rho_1\tilde\rho)\\
	K^\mu \tilde V_\mu = 0\\
	\tilde g = \frac s {4 f_1} \left (b \frac{f_1}{\tilde\rho}-\frac{f_2\rho_3}{\rho_1^2}\right)
\end{gather}
and also equations for the 2-forms 
$F_{\mu\nu}\equiv \der_\mu A_\nu-\der_\nu A_\mu$ and $\tilde{G}_{\mu\nu}$
\begin{multline}
	F_{\mu\nu} = -\frac{2}{\rho_3 (f_1^2+f_2^2)}\bigg[-\left(2-\frac{\rho_3^2}{\rho_1^2}\right)\frac{1}{\rho_3}\epsilon_{\mu\nu\rho\sigma}K^\rho L^\sigma +
	\frac{b}{\tilde\rho}\left(K_\mu L_\nu - K_\nu L_\mu\right)+\\
-f_1 \epsilon_{\mu\nu\rho\sigma}K^\rho\der^\sigma\ln (\rho_3\tilde \rho) -f_2\big( K_\mu \der_\nu \ln(\rho_3\tilde\rho)-K_\nu \der_\mu \ln(\rho_3\tilde\rho)\big)+\\
+4 f_1 \big(K_\mu \tilde V_\nu-K_\nu \tilde V_\mu\big)+4 f_2 \epsilon_{\mu\nu\rho\sigma}K^\sigma \tilde V^\rho\bigg]\label{fmunu}
\end{multline}
\begin{multline}
	\tilde G_{\mu\nu} = -\frac{1}{2(f_1^2+f_2^2)}\bigg[\left(\frac{b}{2 \tilde\rho}-\tilde g s\right)\big(f_1(K_\mu\der_\nu\ln\tilde\rho-K_\nu\der_\mu\ln\tilde\rho) +f_2\epsilon_{\mu\nu\rho\sigma}K^\rho\der^\sigma \ln \tilde\rho\big)+\\
-f_2\left(K_\mu\tilde V_\nu -K_\nu\tilde V_\mu\right)+f_1 \epsilon_{\mu\nu\rho\sigma}K^\rho\tilde V^\sigma\bigg]\label{Gmunu}
\end{multline}
\subsection{Differential relations}
We can use \eqref{mu} to prove the following relations
\begin{gather}
	\tilde\nabla_\mu K_\nu = 4 \left(\tilde G_{\mu\nu}f_1+G_{\mu\nu}f_2\right)- \frac {\rho_3} 2 F_{\mu\nu} f_2 s  +2 \epsilon_{\mu\nu\rho\sigma}\tilde V^\rho K^\sigma s- 2 \tilde{g} Y_{\mu\nu} s\label{dK}\\
\der_\mu \ln f_1 = \der_\mu \ln \tilde \rho\\
\der_\mu \left(\frac{f_2}{\rho_3}\right)=F_{\mu\nu}K^\nu s\,.
\end{gather}
The first equation says that $K^\mu\der_\mu$ is a Killing vector for $ g_{\mu\nu}$. 
We make the natural gauge choice
\begin{equation}
	K^\mu \der_\mu = \der_t\,.
\end{equation}

The second equation can be easily integrated to give, with a suitable choice of  
constant of integration
\begin{equation}
	f_1 = \tilde \rho\,.
\end{equation}
Note further that as a consequence of these equations and of the Bianchi identity
\begin{equation}
	\tilde V = \frac 12 \frac{1}{\rho_3 \tilde\rho^3}\dd (\tilde{g} \rho_1^2 \tilde\rho^3)\, ,
\end{equation}
that $F_{\mu\nu}$ is $t$ independent and we can make a gauge choice for $A_\mu$ such that $\der_t A_\mu = 0$.  Integrating the equation for $f_2$ we obtain
\begin{equation}
	f_2 = \rho_3 (c+A_t s)\,.
\end{equation}
We define the coordinate
\begin{equation}
	y \equiv \rho_1\tilde\rho
\end{equation}
and thus
\begin{equation}
	L_\mu \dd x^\mu = -\frac{\rho_1}{\rho_3}\dd y\,.
\end{equation}
Since $K\cdot L =0$ there is no cross term $g_{ty}$ in the metric. 
We can additionally make a coordinate choice such that there are also no $g_{yi}$ cross terms. 
We have thus reduced our Ansatz for the four dimensional part of the metric to the following
\begin{equation}
	\dd s^2 = - h^{-2} (\dd t + V_1 \dd x^1 + V_2 \dd x^2)^2 + h^2 \frac{\rho_1^2}{\rho_3^2}\tilde h_{ij}\dd x^i\dd x^j +  h^2 \frac{\rho_1^2}{\rho_3^2}\dd y^2\,.
\end{equation}
Note that
\begin{equation}
	h^{-2} = f_1^2 + f_2^2\,.
\end{equation}
For convenience we set
\begin{equation}
	A_y= 0\,.
\end{equation}
All the entries in the metric and in the 5-form are parametrised by a set of functions that we 
can distinguish on the basis of their transformation properties in the $\{x^1,x^2\}$ plane.
\begin{center}
\begin{tabular}{|c|c|c|}
\hline	
Scalars & Vectors & Symmetric Tensor  \\ 
$\rho_1,\rho_3,\tilde\rho , A_t$ & $V_i,A_i$ & $\tilde h_{ij}$\\
\hline
\end{tabular}
\end{center}
Recalling that the scalars are subject to the constraint
\begin{equation}
	y= \rho_1 \tilde\rho\,.
\end{equation}
>From now on we will assume for definiteness that $s=1$.
\subsection{Specifying the spinor}
Due to our gauge choice we have
\begin{gather}
	K^0 = e^0_{\phantom 0 t} = h^{-1} \quad \then \quad \eps^\dagger \eps = h^{-1}\\
	L_3 = L_y E^y_{\phantom y 3} = - \frac{\rho_1}{\rho_3} \frac{\rho_3}{\rho_1}h^{-1} = - h^{-1}\,.\\
\end{gather}
We thus have:
\begin{equation}\label{firstproj}
	\frac{\eps^\dagger \gamma_0 \gamma_5 \gamma_3\eps}{\eps^\dagger\eps} = -1 \quad\then\quad \ii \gamma_1\gamma_2\eps =-\eps\,. 
\end{equation}
We can now take the sum of equations \eqref{at} and \eqref{a12} divided by, respectively, $\tilde\rho$ and $\rho_1$  from which we obtain
\begin{equation}
	\big(\sqrt{1+\e^{-2G}} \gamma_3 \hat \sigma_1 + \ii \gamma_5 \e^{-G} - 1 \big)\eps = 0
\end{equation}
where $\e^{-G} \equiv \frac{f_1}{f_2}$. The solution to this equation is given by
\begin{equation}\label{secondproj}
	\eps = \e^{\ii \delta \gamma_5 \gamma_3 \hat\sigma_1} \eps_1\qquad \gamma_3\hat\sigma_1 \eps_1 = \eps_1
\end{equation}
with $\sinh (2\delta) = \e^{-G}$. The normalization $h^{-1} = \eps^\dagger\eps$ implies $\eps_1 = f_2^{1/2} \eps_0$ with $\eps_0^\dagger\eps_0 = 1$.
These conditions are enough to satisfy all the algebraic equations \eqref{a12},\eqref{a3},\eqref{at}.

Due to the three projectors \eqref{chiralproj},\eqref{firstproj},\eqref{secondproj} and the conditions on the $\hat \chi,\tilde\chi$ spinors, the solution space of the Killing spinor equation is two dimensional and complex. 

We will now use the differential equations \eqref{mu} and the Bianchi identities \eqref{bianchiGVA}-\eqref{lastbianchi} to express the unknown vectors and tensors in terms of the scalars.
\subsection{The spacetime metric and the gauge field A}
We define three new bilinears
\begin{equation}
\begin{split}
	\omega_\mu &= \eps^t \gamma_2 \gamma_\mu\eps\\
W^{1,2}_{\mu\nu} &= \eps^t \gamma_2 \gamma_{\mu}\gamma_{\nu}\hat\sigma_{1,2}\eps\,.
\end{split}
\end{equation}
Using \eqref{mu} we can derive

\begin{equation}
	 \der_\mu\omega_\nu-\der_\nu\omega_\mu = -\ii \frac{\rho_3}{2}F_\mu^{\phantom\mu \rho}W^2_{\nu\rho}-2\ii (A_\mu \omega_\nu-A_\nu\omega_\mu)+ 4\epsilon_{\mu\nu\rho\sigma}\tilde V^\rho \omega^\sigma - 4 \tilde g W^1_{\mu\nu} \,.\label{domega}
\end{equation}
We note that
\begin{equation}
	\omega_\mu\dd x^\mu = -\frac{\rho_1}{\rho_3} (\tilde e^1_{\phantom 1 j}+ \ii \tilde e^2_{\phantom 2 j} )\dd x^j\equiv - \frac{\rho_1}{\rho_3} \tilde e^z_{\phantom z j}\dd x^j
\end{equation}
where $\tilde e^k_{\phantom k j}$ is a 2-bein for the metric $\tilde h_{ij}$. Thus, from \eqref{domega} we can get an equation involving $\dd \tilde e^k$.
Singling out the $y$ dependence using the $(y,x^i)$ component of \eqref{domega}
\begin{equation}
\der_y \tilde e^z_{\phantom i j}=-2 \frac{h^2}{\rho_3\rho_1}\left[\frac{\tilde\rho}{\rho_3}(\rho_3^3-\rho_1^2)+\frac{f_2}{\tilde\rho}(f_2\rho_3-b \rho_1^2)\right]\, \tilde e^z_{\phantom i j}\equiv D \tilde e^z_{\phantom i j}\,.
\end{equation}
With a further $y$ independent coordinate transformation we can put $\tilde h_{ij}$ in diagonal form. We introduce a conformal factor $T$ and set
\begin{equation}
	\tilde e^i_{\phantom i j} = T \delta^i_{\phantom i j}\qquad \der_y T = D T
\end{equation}
Looking at the $(x^1,x^2)$ component we can establish a relation between the remaining derivatives of $T$ and the connection $A_i$
\begin{equation}
	A_i = (A_t+b-c) V_i - \frac 12 \epsilon_{ij} \der_j \ln T\,.
\end{equation}
The constant $c$ can be absorbed into a gauge transformation and we will set
\begin{equation}
	b=c=1\,.
\end{equation}
The right hand side of the $\{t,x^i\}$ component of equation \eqref{domega} is proportional to $b-c$ and thus also this equation is consistent with our gauge choice.\\
We have now an expression for $A_i$ by which we may calculate the components of $F_{\mu\nu}$. This $F_{\mu\nu}$ must be equal to the one obtained in \eqref{fmunu}. The contraction with $K^\mu$ is trivial. The $F_{yi}$ components give the constraint
\begin{equation}
	b=c
\end{equation}
which is solved by our gauge choice. The $F_{12}$ component gives an equation for $(\der_1^2+\der_2^2) T$ that we will discuss later.

We have thus reduced our set of unknowns to five scalars and one 2-vector. Two scalars are constrained by the relations $y=\rho_1 \tilde\rho$ and so we have just four independent scalars and one 2-vector.

\begin{center}
\begin{tabular}{|c|c|}
\hline	
Scalars & Vector \\ 
$\rho_1,\rho_3,\tilde\rho , A_t,T$ & $V_i$\\
\hline
\end{tabular}
\end{center}
We have reduced the four dimensional metric to the form
\begin{equation}
	\dd s^2 = - h^{-2} (\dd t + V_i \dd x^i)^2 + h^2 \frac{\rho_1^2}{\rho_3^2}(T^2\delta_{ij}\dd x^i\dd x^j + \dd y^2)\,.
\end{equation}

To simplify the final equations we now express the 4 functions $\rho_1,\rho_3,\tilde\rho, A_t$ in terms of three independent functions that we will call $m,n,p$ are defined by
\begin{equation}	
\begin{array}{ll}
 \rho_1^4 = y^4 \frac{m p+ n^2}{m}& \rho_3^4 = \frac{p^2}{m(mp+n^2)} \\ 
 \tilde\rho^4 = \frac{m}{mp+n^2}& A_t = \frac{n-p}{p} 
\end{array}
\end{equation}
With these definitions we have
\begin{equation}
	D = 2y(n+m-y^{-2})\,.
\end{equation}

With some effort it can be shown that all the equations on the spinor $\eps$ are now solved. 

As noted in Section B.6 the space of solutions to the Killing spinor equation is 2-dimensional and complex, thus our backgrounds preserve 4 of the 32 real supersymmetries of the theory. The existence of the Killing spinors guarantees that the full Einstein equations are satisfied provided that integrability conditions and the Bianchi identities for the Ramond-Ramond 5-form are satisfied. Let us now investigate what the consequence of these final constraints are.
\subsection{Differential Equations}
We will first establish a relation between the vector $V_i$ and the various scalar functions. The equation \eqref{dK} is an equation for $\dd K$ with
\begin{equation}
	K= - h^{-2} (\dd t + V_i \dd x^i)\,.
\end{equation}
We can extract from this equation an expression for $\dd V$:
\begin{equation}
	\dd V = -y \star_3 [ \dd n + (n D +2y m (n-p) + 2n/y)\dd y]
\end{equation}
where by $\star_3$ we mean the Hodge dual in the three dimensional diagonal metric
\begin{equation}
	\dd s_3^2 = T^2\delta_{ij}\dd x^i\dd x^j + \dd y^2\,.
\end{equation}
Returning to the Bianchi identities
\begin{gather}
	\dd \big(\tilde G \tilde \rho^3 - \tilde V \wedge A \rho_3\tilde\rho^3\big) = 0\label{bianchiGVA2}\\
	\tilde V = \frac 12 \frac{1}{\rho_3 \tilde\rho^3}\dd (\tilde{g} \rho_1^2 \tilde\rho^3)\label{bianchitildeV2}\\
	\dd \big( G \rho_1^2 \rho_3\big) = 0\label{bianchiG2}\\
	\dd \big( G \rho_1^2 \rho_3 \wedge A + \star_4 \tilde V \big)-2\star_4 \tilde{g} = 0\,.\label{lastbianchi2}
\end{gather}
Substituting in the first equation $\tilde V$ as obtained from the second equation we find
\begin{equation}
	\dd \big(\tilde G \tilde \rho^3 - \frac12 \tilde{g} \rho_1^2 \tilde\rho^3  F\big) = 0\,.
\end{equation}
We may thus set locally
\begin{equation}
\begin{split}
&\dd \tilde B = \tilde G \tilde \rho^3 - \frac12 \tilde{g} \rho_1^2 \tilde\rho^3  F \\ 
&\tilde B = \tilde B_t (\dd t + V) + \hat{\tilde B}\\
&\dd B =  G \rho_1^2 \rho_3\\
&B = B_t (\dd t +V) + \hat B\,.
\end{split}
\end{equation}
The algebraic equation \eqref{Gmunu} for $\tilde G_{\mu\nu}$ and for its dual for $G_{\mu\nu}$ give rise to four new relations
\begin{equation}
\begin{split}
 & \tilde B_t = -\frac{1}{16} y^2\, \frac{n-1/y^2}{p}\\
& \dd \hat{\tilde B} = -\frac1{16} y^3\star_3 [\dd m + 2 m D]\\
& B_t = -\frac 1 {16} y^2\,\frac{n}{m}\\
& \dd \hat B = \frac 1 {16}y^3\star_3  [\dd p + 4yn(p-n)\dd y]\,.
\end{split}
\end{equation}
We need to impose the three equations
\begin{equation}
\left\{ \begin{array}{l}
\dd \dd V = 0\\
\dd\dd \hat{\tilde B}=0\\
\dd\dd \hat B =0
\end{array}
\right.\,.
\end{equation}
The last Bianchi identity \eqref{lastbianchi2} is implied by these three. In addition to these equations we have also
\begin{equation}
	\der_y \ln T = D
\end{equation}
which together with the previous ones can be used to see that also the consistency equation for $F_{12}$ is satisfied.

We have thus a set of 4 equations for 4 unknowns: $m,n,p,T$. The equations are defined on the half space
\begin{equation}
	(x^1,x^2,y>0)
\end{equation}
and are quite complicated being a set of coupled non-linear second order elliptic differential equations. 
\begin{equation}
\begin{split}
&y^3 (\der_1^2+\der_2^2) n + \der_y \left(y^3 T^2 \der_y n\right) + y^2\der_y\big[ T^2 \big(y D n +2y^2m(n-p)\big)\big]+4 y^2 D T^2 n=0\\
&y^3 (\der_1^2+\der_2^2) m + \der_y \left(y^3 T^2 \der_y m\right) + \der_y \left( y^3 T^2 2m D\right)=0\\
&y^3 (\der_1^2+\der_2^2) p + \der_y \left(y^3 T^2 \der_y p\right) + \der_y \big[ y^3 T^2 4n y(n-p)\big]=0
\end{split}
\end{equation}
\section{Killing vectors and the Kaluza Klein Ansatz}
In this appendix we present a geometrical interpretation of the bilinears that we constructed and that we used in Appendix B

Assume we have a fibration of a group manifold over some $d$ dimensional base manifold with metric
\begin{equation}
	\dd s^2 = \tilde g_{\mu\nu}(x)\dd x^\mu \dd x^\nu + \beta_{ab}(x) \left( \hat e^a (y) - A^a_{\mu}(x) \dd x^\mu \right)\left( \hat e^b (y) + A^b_{\mu}(x) \dd x^\mu \right)
\end{equation}
where $\hat e^a$ is a basis of left-invariant one forms on the group manifold.

We define
\begin{equation}
	\kappa =  K^\mu\der_\mu + \alpha(x)^a \hat E_a\,.
\end{equation}

We recall that given any covariant 2-tensor $a$ and three vector $W,V_1,V_2$ the Lie derivative of $a$ is given by
\begin{equation}
	(\mathcal L_{W} a )(V_1,V_2) = W \left( a(V_1,V_2)\right) - a ( [W,V_1],V_2)-a(V_1,[W,V_2])\,.
\end{equation}
Let us calculate $\mathcal L_K g$
\begin{gather*}
	\left( \mathcal L_{\kappa} g \right)(\der_\mu,\der_\nu) = \left( \mathcal L_{K}\tilde g \right)(\der_\mu,\der_\nu)+ K^\rho \der_\rho \left(\beta_{ab} A^a_\mu A^b_{\nu}\right)-\der_\mu \alpha^a \beta_{ab}A^b_{\nu}-\der_\nu \alpha^a\beta_{ab} A^b_{\mu}\\
\left( \mathcal L_{\kappa} g \right)(\hat E_a,\hat E_b) = K^\rho \der_\rho \beta_{ab}+\left( \mathcal L_{\alpha} \hat g(x) \right)(\hat E_a,\hat E_b)\\
\left( \mathcal L_{\kappa} g \right)(\der_\mu,\hat E_a) =K^\rho\der_\rho (-\beta_{ab} A^b_{\mu})+\beta_{ab}\der_\mu\alpha^b+\beta_{cd} \alpha^b A^d_\mu f_{ba}^c
\end{gather*}
where $\hat g = \beta_{ab}\hat e^a \hat e^b$ and $f_{ba}^c$ the structure constants of the group.

When $K =0, \beta_{ab} = k_{ab}$ with $k_{ab}$ the Killing form of the group and so we obtain the non abelian Kaluza Klein setup.

Assume for the moment that $K$ is a Killing vector of $\tilde g$ and $\alpha^a \hat E_a$ is a Killing vector of $\hat g$, what are the conditions on $\alpha,\beta_{ab},A^a_\mu$ such that $K$ is a Killing vector for the whole metric?
This is easily seen from our previous equations
\begin{gather}
	K (\beta_{ab}) = 0\\
	\der_\mu \alpha^a =  K(A^a_\mu) -f_{bc}^a \alpha^b A^c_{\mu}\,.
\end{gather}

We can now specialise to our setting. The group manifold is $SU(2)\times SU(2)$. We can define the ten dimensional vector
\begin{equation}
\begin{split}
	&\kappa^M  \der_M = \bar\psi\Gamma^M\psi\der_M =\\
	&K^\mu\der_\mu + \left( A_m K^m  - \frac{f_2}{\rho_3} s\right) \Sigma^{\hat i}_{\phantom i \hat 3} \der_{\hat i} + \frac{f_1}{\tilde\rho} \tilde \chi^\dagger \sigma^{\tilde a}\tilde\chi \Sigma^{\tilde i}_{\phantom i \tilde a}\der_{\tilde i}
\end{split}
\end{equation}
where we have chosen the normalization $\chi^\dagger\chi =  \tilde \chi^\dagger\chi = 1$. $\kappa$ is a Killing vector and it is null \cite{Hackett-Jones:2004yi}. Since it is null we have
\begin{equation}
	K^2 =K_\mu \tilde g^{\mu\nu} K^\nu= -f_1^2-f_2^2
\end{equation}
which was previously seen as consequence of Fierz rearrangements and whereas here we can see its geometrical origin. From the equation on $\nabla_\mu K_\nu$ we know that $K$ is a Killing vector for $\tilde g_{\mu\nu}$, and moreover, due to the Killing equation on $\tilde \chi$ and the properties of the Ansatz, we have that
\begin{equation}
	 \Sigma^{\hat i}_{\phantom i \hat 3}\qquad , \qquad \tilde \chi^\dagger \sigma^{\tilde a}\tilde\chi \Sigma^{\tilde i}_{\phantom i \tilde a}\der_{\tilde i}
\end{equation}
are Killing vector of the group manifolds. We thus conclude
\begin{gather}
K(\rho_1) = K (\rho_3) = K (\tilde \rho) = 0\\
	\der_\mu \left(\frac{f_1}{\tilde  \rho} \right) = 0\\
	\der_\mu \left( A_\nu K^\nu  - \frac{f_2}{\rho_3} s\right) = K(A_\mu)
\end{gather}

The second one can be written in the form we already encountered earlier
\begin{equation}
	\der_\mu \left( \frac{f_2}{\rho_3}\right) = F_{\mu\nu} K^\nu s\,.
\end{equation}

We have thus clarified the geometrical origin of the relations between $f_1,f_2$ and the metric entries.
\providecommand{\href}[2]{#2}\begingroup\raggedright\endgroup

% \bibliographystyle{JHEP}
% \bibliography{biblioLLMcompany}
\end{document}